\documentclass[prd,onecolumn,nofootinbib,aps,tightenlines,preprintnumbers,notitlepage,longbibliography,superscriptaddress,12pt]{aastex631}

\usepackage{color}
\usepackage{tikz}
\usepackage[caption=false]{subfig}
\usepackage{hyperref}
\usepackage{multirow}
\usepackage{graphicx}
\usepackage{dcolumn}
\usepackage{bm}
\usepackage[normalem]{ulem}
\usepackage{epstopdf}
\usepackage{tabularx}
\usepackage{suffix}
\usepackage{mathtools}

\usepackage{xcolor}

\usepackage{graphicx}
\usepackage{dcolumn}
\usepackage{bm}
\usepackage[normalem]{ulem}
\usepackage{xcolor}

\graphicspath{ {figures/} }

\usepackage{epstopdf}
\newcommand{\be}{\begin{eqnarray}}

\newcommand{\ee}{\end{eqnarray}}

\def\baseline{{{baseline}}}
\def\unlensed{{{unlensed}}}

\def\spinup{\partial\kern-0.3em\raise0.42ex\hbox{\tiny\textbackslash}}
\def\spindown{\overline{\partial\kern-0.3em\raise0.42ex\hbox{\tiny\textbackslash}}}

\newcommand{\dd}{{\rm d}}

\definecolor{purpleCIB}{rgb}{0.62, 0.0, 0.77}
\definecolor{salmonpink}{rgb}{1.0, 0.57, 0.64}

\def\p{\parallel}

\def\k{\boldsymbol{k}}

\def\d{{\rm d}}

\def\nhat{\hat{\boldsymbol{n}}}

\usepackage{suffix}
\usepackage{mathtools}

\DeclarePairedDelimiterX\MeijerM[3]{\lparen}{\rparen}%
{\begin{smallmatrix}#1 \\ #2\end{smallmatrix}\delimsize\vert\,#3}

\newcommand\MeijerG[8][]{%
  G^{\,#2,#3}_{#4,#5}\MeijerM[#1]{#6}{#7}{#8}}

\WithSuffix\newcommand\MeijerG*[7]{%
  G^{\,#1,#2}_{#3,#4}\MeijerM*{#5}{#6}{#7}}

\definecolor{dodgerblue}{rgb}{0.12, 0.56, 1.0}
\definecolor{orange}{rgb}{1.0, 0.5, 0.0}

\newcommand{\jhu}{William H. Miller III Department of Physics and Astronomy, Johns Hopkins University, Baltimore, MD 21218, USA}

\newcommand{\perimeter}{Perimeter Institute for Theoretical Physics, 31 Caroline St N, Waterloo, ON N2L 2Y5, Canada}

\newcommand{\usc}{Physics \& Astronomy Department, University of Southern California, Los Angeles, California, 90089-0484}

\begin{document}

\author{Selim~C.~Hotinli}
\affiliation{\perimeter}
\affiliation{\jhu}

\author{Elena~Pierpaoli}
\affiliation{\usc}

\title{On the Detectability of the Moving Lens Signal in CMB Experiments}

\date{\today}


\begin{abstract}
    
Upcoming cosmic microwave background (CMB) experiments are expected to detect new signals probing interaction of CMB photons with intervening large-scale structure.  Among these the moving-lens effect, the CMB temperature anisotropy induced by cosmological structures moving transverse to our line of sight, is anticipated to be measured to high significance in the near future. In this paper, we investigate two possible strategies for the detection of this signal: pairwise transverse-velocity estimation and oriented stacking. We expand on previous studies by including in the analysis realistic simulations of competing signals and foregrounds.
We confirm that the moving lens effect can be detected at  $\ge 10\sigma$ level  by a combination of CMB-S4 and LSST surveys.
We show that the limiting factors in the detection depend on the strategy:
for the stacking analysis, correlated extragalactic foregrounds, namely the 
cosmic infrared background and thermal Sunyaev Zel'dovich effect,  play the most important role. The addition of foregrounds make the signal-to-noise ratio be most influenced by large and nearby objects. As for  the pairwise detection, halo lensing and pair number counts  are the main issues.
In light of our findings, we elaborate on possible strategies to improve the analysis approach for the moving lens detection with upcoming experiments. 
We also deliver to the community all the simulations and tools we developed for this study.

\end{abstract}


\section{Introduction} \label{sec:intro}

The anisotropies of cosmic microwave background (CMB) provide a wealth of information and has become the backbone of modern cosmological inference. Improvements in the sensitivity of CMB experiments are now opening the window to high-precision measurements of small-scale CMB anisotropies sourced by the interactions of CMB photons with the intervening matter, providing promising new probes of cosmology and astrophysics. These anisotropies include scattering of CMB photons off energetic electrons (Sunyaev Zel'dovich effects)~\citep{1969Ap&SS...4..301Z, 1970A&A.....5...84Z, 1980ARA&A..18..537S, 1972CoASP...4..173S, Sazonov:1999zp}, weak gravitational lensing~\citep[see e.g.][for a review]{Lewis:2006fu} and the integrated Sachs-Wolfe (ISW) effects~\citep{1967ApJ...147...73S}. In this work, we focus on  a particular type of ISW: the moving-lens effect \citep[otherwise also known as Birkinshaw-Gull, hereafter ML ][]{1983Natur.302..315B,Gurvitz:1986ab}. 

The ML effect is  sourced by the time variation of the gravitational potentials due to peculiar velocities of halos in the direction perpendicular to our line of sight. 
The expected signal is small compared to other CMB foregrounds (typically of the order of $\simeq 0.1 - 0.01 \mu \rm{K}$) and has the same frequency dependence as  the primary CMB.  
However, this is one of the two {physical effects} that potentially allow us to determine transverse velocities of large-scale structure from CMB \citep[the other being the polarized kinetic Sunyaev-Zel'dovich effect][]{Hotinli:2022wbk}, and it is therefore of great interest to pursue.

Velocity fields are powerful probes of the growth of large-scale structure, and can be related to the underlying matter perturbations on large scales. So far, only the radial velocity field has been inferred from CMB observations by measurements of the kinetic Sunyaev-Zel'dovich (kSZ) effect, which has important cosmological applications~\citep[e.g.][]{Munchmeyer:2018eey,AnilKumar:2022flx,Hotinli:2022jna,Hotinli:2019wdp}. Specifically, recent CMB experiments allowed the detection of the pairwise radial velocity: a statistic sensitive to the average gravitational attraction of pairs of galaxies at a given comoving distance~\citep{2012PhRvL.109d1101H, Planck:2015ywj, DeBernardis:2016pdv, DES:2016umt}. These measurements used a pairwise-velocity estimator based on the observed radial velocity \citep{Ferreira:1998id}, similar in spirit to the one used in this work~\citep{Yasini:2018rrl}, which is based on transverse velocities~\citep{Hotinli:2018yyc}. 

While the kSZ is expected to be measured at very high significance with future CMB surveys~\citep{Smith:2018bpn,Cayuso:2021ljq}, preliminary studies suggest that transverse velocities will also be detected~\citep[e.g.][]{Hotinli:2018yyc,Yasini:2018rrl,Hotinli:2020ntd}, albeit to lower significance.
Nevertheless, this supplemental information will indeed improve the determination of the cosmological 3-dimensional velocity field.
Furthermore, the ML effect is subject to different systematics than the kSZ.
The cosmological information available from the kSZ effect is limited by the degeneracy of this signal with the optical depth of galaxies \citep[see e.g.][for a detailed discussion]{Smith:2018bpn}. 
While mitigating strategies have been proposed \cite[e.g.][]{Madhavacheril:2019buy}, it is not clear whether they can be readily implemented in the near future.
The measurement of the transverse velocities hence have the potential of improving the information content of the reconstructed 3-velocity field, as well as potentially breaking degeneracies suffered by kSZ measurements~\citep[see e.g][]{Hotinli:2021hih}. 

In this work, 
we expand on previous studies on the ML detectability ~\citep{Hotinli:2018yyc, Yasini:2018rrl, Hotinli:2020ntd} by considering a more realistic setting for the analysis of upcoming surveys like the Simons Observatory~\citep{Ade:2018sbj,Abitbol:2019nhf} and CMB-S4~\citep{Abazajian:2016yjj,CMB-S4:2022ght}, together with the  Vera Rubin Observatory~(LSST)~\citep{2009arXiv0912.0201L} and DESI~\citep{DESI:2016fyo}.

We perform a map-based analysis of the detectability of the ML signal with two different methods: oriented stacking of CMB patches around DM halos and pairwise-velocity reconstruction, discussing the  advantages and limitations for  each of them, when standard component separation methods are applied to the observed maps.
For the first time, 
we use in our analysis  realistic simulations of correlated and non-Gaussian CMB foregrounds, including thermal and kinetic Sunyaev Zel'dovich effects (tSZ and kSZ), the cosmic infrared background (CIB), halo lensing, radio point sources, as well as the ML effect from dark matter (DM) halos of all masses (see Fig.~\ref{fig:mixedmaps} for a display of these effects).
As pointed out in \cite{Hotinli:2023ywh},  the ML measurement from stacking is challenged by physical effects that are correlated with ML.  In general,  foregrounds  will dictate which halo masses and redshifts are most relevant for the detection,  and ultimately what can be measured with each detection strategy. 
In our analysis, we also investigate the relevance of photo-$z$ errors and uncertainly in the halo mass estimation.

Throughout this paper, {we use the flat $\Lambda$CDM cosmology, with parameters satisfying $(\Omega_m,\Omega_b,\sigma_8,n_s
h,\tau) = (0.31, 0.049, 0.81, 0.965, 0.68, 0.055)$ consistent with Planck 2018~\citep{Planck:2018vyg} {and the \texttt{websky} simulation~\citep{Stein:2020its}}. Unless otherwise stated, we define the halo mass to be $M_{200c}$, corresponding to the mass contained within a radius $r_{200}$ inside of which the mean interior {mass} density is 200 times the critical density $\rho_{\rm crit}$.} 

This paper is organized as follows. In Sec.~\ref{sec:ML} we introduce the ML signal. In Sec.~\ref{sec:SIMS} we introduce the dark matter halo catalog and the extra-galactic CMB simulations we use in our analysis. Section~\ref{sec:exp} describes our choices for the experimental specification we consider. In Sec.~\ref{sec:detect} we analyse the performance of two methods for detecting the ML effect with upcoming CMB and galaxy data. Section~\ref{sec:discussion} is dedicated to our discussion and conclusions.

\section{The moving lens effect} \label{sec:ML}

Gravitational potentials that evolve in time induce a black-body temperature modulation on the CMB known as the integrated Sachs-Wolfe (ISW) effect.  The induced temperature anisotropy in a given direction $\nhat$ has the form 
\begin{equation}
\Theta_{\rm ISW}(\nhat)=-\frac{2}{c^3}\int\dd \chi\, \dot{\Phi}(\chi\nhat)\,,
\end{equation}
where $\Phi$ is the gravitational potential. {Here, $\Theta(\nhat)=T(\nhat)/T_{\rm CMB}$ is the fractional CMB temperature where $T_{\rm CMB}$ is the sky-averaged mean CMB temperature.}  
The ISW effect can be sourced by peculiar transverse velocities of cosmological structure where the effect on the CMB satisfies
\begin{equation}\label{eq:ML_theta1}
\Theta_{\rm ML}(\nhat)=-\frac{2}{c^3}\int\dd \chi\,\boldsymbol{{v}}_\perp(\chi\nhat)\cdot\nabla_{\!\perp}\Phi(\chi\nhat)\,,
\end{equation}
where $\nabla_{\!\perp}$ is the gradient on the 2-sphere. {This signal is referred to as ML effect in cosmology literature due to the analogy with gravitational lensing, since, for the observer on the rest frame of the halo, the effect is equivalent to lensing.} 

{Note that the ISW effect on small scales is also sourced by the non-linear collapse of matter around galaxies and clusters, which induces a radially-symmetric imprint on the CMB around halos, known as the Rees-Sciama (RS) effect. Here, we are interested in measuring the dipolar signature from the moving lens effect aligned with transverse velocities of cosmological structure, and our estimators will be insensitive to the RS effect. Prospects to detect the RS signal has been recently discussed in~\cite{Ferraro:2022twg}}. 

The ML effect sourced by the bulk transverse velocity of DM halos, {assumed nearly constant within the range where the integrand in Eq.~\eqref{eq:ML_theta1} in non-vanishing}, can be further simplified to take the form
\be
\Theta_{\rm ML}^{(h)}(\nhat)= \frac{\boldsymbol{{v}}_{\perp}(\chi\nhat)}{c}\cdot \boldsymbol{\beta}_{h}(\chi\nhat)\,,
\ee
where $\boldsymbol{\beta}_{h}=-2\boldsymbol{\nabla}\!\int\dd \chi\,({1}/{\chi})\Phi_{h}(\chi\nhat)/c^2$, is the lensing deflection vector, {$\boldsymbol{\nabla}$ is the three-dimensional angular derivative, satisfying $\boldsymbol{\nabla}_\perp=(1/\chi)\boldsymbol{\nabla}$,} and $\Phi_{h}$ is the gravitational lensing {only} potential of a halo. The top left two panels in Fig.~\ref{fig:mixedmaps} demonstrate the ML effect in a $2^o\times2^o$ patch. The right panel includes the signal from {the most massive} halos. The dipolar ML signature aligned with the bulk velocity can be seen around the central halo in that panel. Furthermore, the ML signal can be seen to extend beyond the virial radius of the halo, which is around 10 arcminutes in angular size here. The left panel corresponds to the total signal from the ML effect including all halos in the \texttt{webksy} simulation in that patch. The otherwise apparent dipolar pattern from the individual high mass halo can be seen to be distorted by the ML signal from other (smaller-mass and higher-redshift) halos. Other  panels demonstrate the effect of various CMB foregrounds, which we describe in the next paragraph. 

Because the ML effect has a black-body frequency dependence, its signature will be imprinted in the CMB map that gets reconstructed from multi-frequency observations. We expect the same map to also contain the other effects which follow a  black-body spectrum, namely the  kSZ and halo lensing. Moreover, some residual signal from other foregrounds can remain in the CMB reconstructed map. 
As  Fig.~\ref{fig:mixedmaps} shows, the amplitude of the ML signal is very small, and it  is subdominant with respect to other competing signals as well as to the residual foregrounds.  The ML signal is likely to be detected statistically for an ensemble of objects as opposed to for individual ones. This paper studies in detail the detectability of the ML while considering ensembles of halos, and considering the challenges posed by the other sky signals.

\section{Dark matter halo catalog and extragalactic CMB simulations} \label{sec:SIMS}

\begin{figure}[t]
\centering
\includegraphics[width=0.8\linewidth]{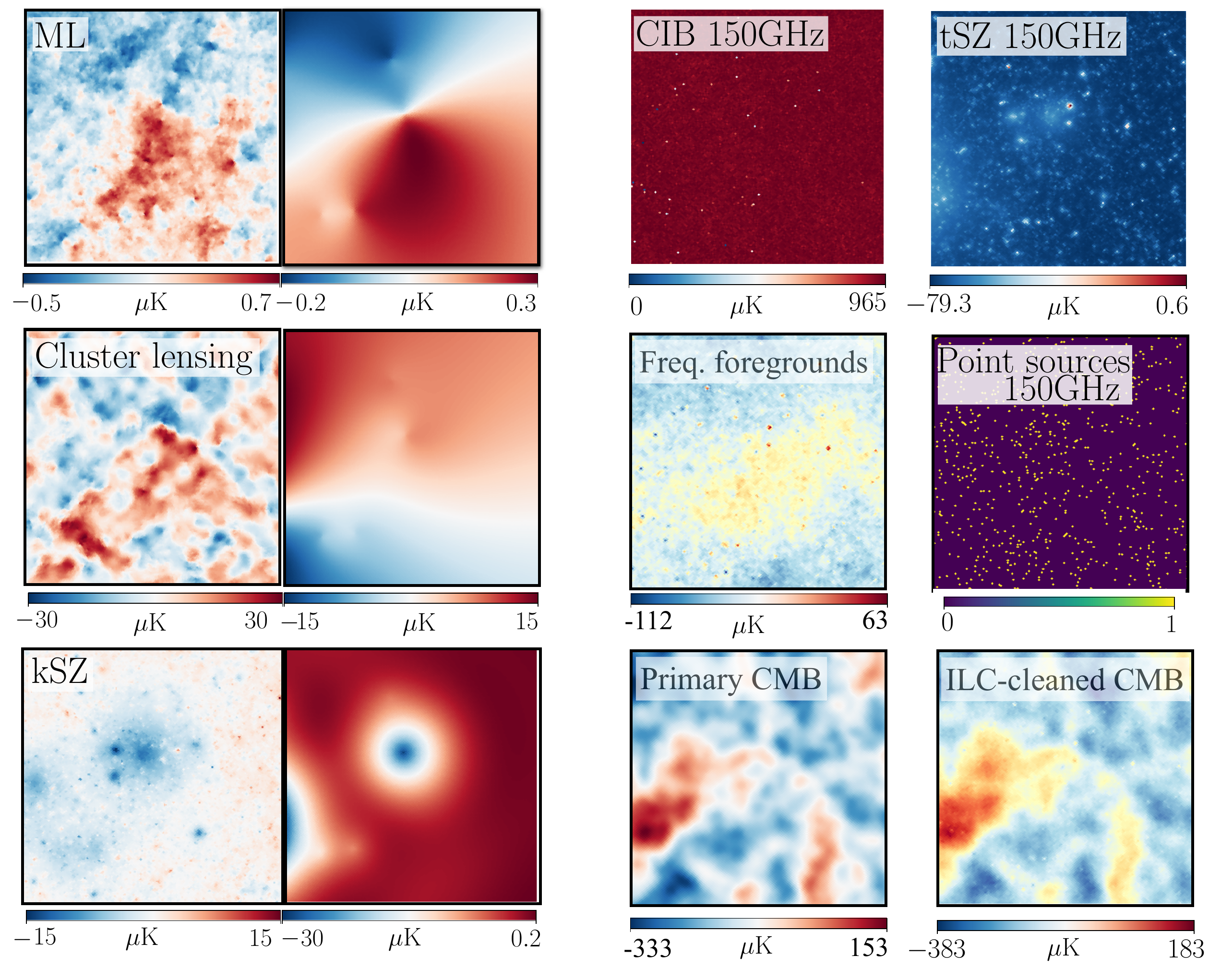}
\vspace*{-0.3cm}
\caption{The panels show $ 2^o\times2^o $ patches of the ML signal, the CMB,  and relevant foregrounds, as computed from the \texttt{websky} simulations in the same area of the sky. The two  top-left panels show the ML signal, the two
central-left  panels show the cluster lensing, while the lower left panels show the kSZ effect.
Within these, the left panels  show the maps of $\sim 8000$ halos of mass $M_h>10^{12}M_\odot$ within $0<z<4.6$; while the right panels include only halos with $M_h\gtrsim10^{14}M_\odot$. 
All the  above-mentioned  effects share the same black-body  frequency dependence of the primary CMB (which is shown in the bottom right).
Other foregrounds have different frequency dependencies.
The top-right panels show the CIB and tSZ signals at 150GHz, while the central-right panel show the signal from point sources at the same frequency, {appropriate for CMB-S4 and based on the \texttt{websky} catalog, provided in \cite{Li:2021ial}.} The bottom-right panel shows the ILC reconstructed map for all signals with a black-body frequency spectrum. The panel marked ``Freq. foregrounds" show the residual level of foreground signal in the ILC {reconstructed map using CMB-S4 level white noise and beam. Note that throughout this study we omit considering the lensing of CMB foregrounds such as CIB and tSZ which might induce further, albeit small, correlations between the LSS and observed CMB maps.}}
\label{fig:mixedmaps}
\end{figure}

Throughout this work we use the \texttt{websky}\footnote{\href{https://mocks.cita.utoronto.ca/data/websky/v0.0/}{mocks.cita.utoronto.ca/data/websky}} DM halo catalog and extragalactic CMB simulations~\citep{Stein:2020its} which share the same cosmological parameters cited in Sec.~\ref{sec:intro}. The \texttt{websky} DM halo catalog is modelled with ellipsoidal collapse dynamics with the corresponding displacement field is modelled with Lagrangian perturbation theory. The catalog spans a redshift interval $0<z<4.6$ over the full-sky spanning a volume of $\sim(600 {\rm Gpc}/h)^3$ and consists of approximately a billion halos of  mass satisfying $M_h>10^{12} M_\odot$.

The halo catalog and the displacement field are also used to generate a range of intensity maps to simulate scattering and lensing effects on the CMB photons. These publicly available maps include the CIB, infrared emission from dusty star forming galaxies; tSZ, inverse Compton scattering of the CMB photons by the energetic free electrons in the intergalactic media; kSZ, Doppler boosting of CMB photons due to Thomson scattering off free electrons; as well as the weak gravitational lensing of the CMB due to intervening large-scale structure. 
In addition to CIB and tSZ, we use maps of radio point sources based on the \texttt{websky} catalog provided in~\citep{Li:2021ial}.
We generate halo lensing from the \texttt{websky} catalog using the \texttt{AstroPaint} code\footnote{\href{https://github.com/syasini/AstroPaint}{github.com/syasini/AstroPaint}}~\citep{Yasini2020}.\footnote{These maps can be found at \hyperlink{https://github.com/selimhotinli/moving_lens}{selimhotinli/moving\_lens}.}

Fig.~\ref{fig:mixedmaps} shows the overall amplitude and morphological shape of these foregrounds. 
{The typical  foregrounds signal is multiple orders of magnitude larger than ML effect.
In this work, we will use the standard Internal Linear Combination (ILC)  component separation method to perform foreground cleaning and recover the ``CMB map", that is the map which contains only the black-body signals. 
As it is visible in the panel titled `Freq. foregrounds' of Fig.~\ref{fig:mixedmaps}, the residual contribution of foregrounds  is still orders of magnitude larger than the ML effect.}

\section{Experiments}
\label{sec:exp}

The anticipated frequency, area coverage, angular resolution and white noise levels matching Simons Observatory~ (SO, \cite{Ade:2018sbj, Abitbol:2019nhf}) and CMB-S4~\citep{Abazajian:2016yjj} are shown on Table~\ref{tab:cmb_specs}.  The measurement of the ML effect through the methods outlined in 
this paper also requires a galaxy catalog with galaxy locations. The galaxy survey specifications we consider here are shown in Table~\ref{tab:lss_specs}. We consider the DESI~\citep{DESI:2016fyo} and LSST~\citep{2009arXiv0912.0201L} as our representative set of galaxy surveys. Specifically, we will combine data from DESI with  SO, and from LSST with CMB-S4.

DESI is an ongoing survey that aims to measure over 30 million spectroscopic galaxy and quasar redshifts over 14000 deg$^2$. The DESI galaxy catalogue could be separated into the Bright Galaxy Sample (BGS), LRG, emission-line galaxy (ELG) sample and the quasar (QSO) sample. BGS is anticipated to have a number density larger than BOSS for $z\lesssim0.4$ and the same is anticipated for DESI LRGs over $0.6\lesssim z\lesssim1$; the ELG sample over $0.6\lesssim z\lesssim1.6$ and the QSO sample over $0.6\lesssim z\lesssim1.8$.

\begin{figure}[t]
\centering
\includegraphics[width=0.48\linewidth]{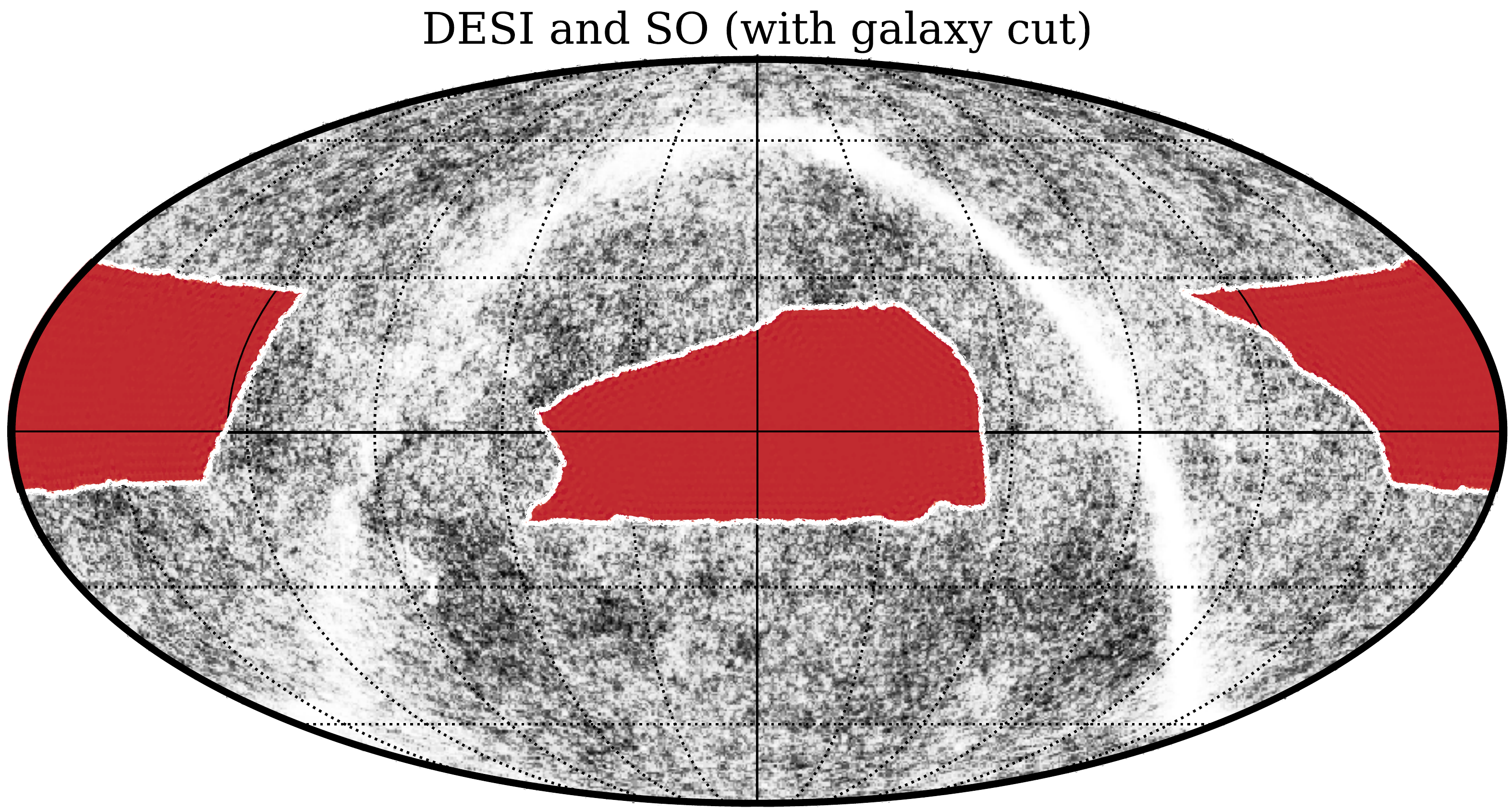}\,\,\includegraphics[width=0.48\linewidth]{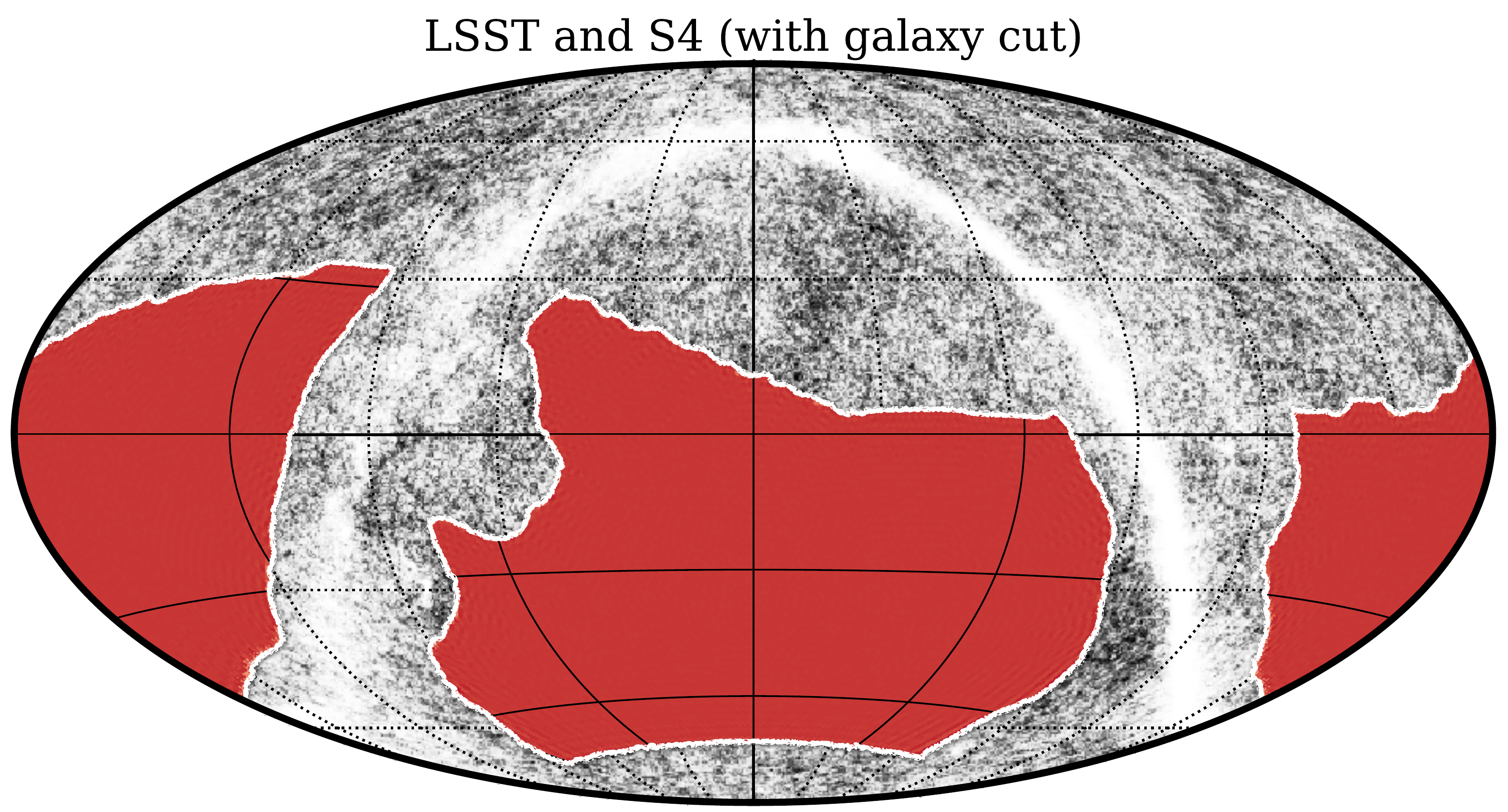}\caption{Joint sky coverage (red) of the upcoming CMB experiments and galaxy surveys. (\textit{Left}) Sky coverage of DESI and Simons Observatory ($f_{\rm sky}\simeq0.20$).  (\textit{Right}) Sky coverage of LSST and CMB-S4 ($f_{\rm sky}\simeq 0.45$).
These maps were generated in part using code available at \href{https://github.com/syasini/cmb-x-galaxy-overlaps}{github/cmb-x-galaxy-overlaps} \citep{Coulton:2021ekh}.}\label{fig:sky_frac}
\end{figure}

Rubin LSST is an ongoing survey that aims to measure up to 2 billion galaxies at the end of its 10 years observation which will cover a sky area around $\sim20000$ deg$^2$. In what follows we assume the redshift error of LSST survey satisfy $\sigma_z/(1+z)=0.03$ as suggested by~\citep{LSSTDarkEnergyScience:2018jkl}. The intersection for the sky coverage  of DESI with SO and LSST with CMB-S4 are shown in Fig.~\ref{fig:sky_frac} ($f_{\rm sky}=0.20$ and $f_{\rm sky}=0.45$, respectively).
For LSST, we approximate the galaxy density of the ``gold'' sample, with $n(z) = n_\text{gal}({z}/{z_0})^2 \exp(-z/z_0)/{2z_0}$ with $n_\text{gal}=40~\text{arcmin}^{-2}$ and $z_0=0.3$ and take the galaxy bias as $b_g(z)=0.95{(1+z)}$. 

While we have galaxy number count predictions for LSST and DESI, \texttt{websky} is a halo catalog.
Large halos may contain more than one galaxy, therefore we cannot  apply  a one-to-one correspondence between halos and galaxy numbers.
We use the Halo occupation distribution model as described in~\cite{2011ApJ...738...45L} to decide how many halos should be considered to be detected in DESI and LSST. The resulting number of halos is a bit (around 10 per cent or less) lower than the number of galaxies that result from table. We verified that the details of the strategy for  halo occupation model is not relevant for any of the analysis  presented in this paper.

\begin{center}
\begin{table*}
\begin{tabular}{| c | l  | c | c | c | c | c |}\hline 
       CMB Experiment & Frequency in GHz = & 40 & 90 & 150 & 220  & Sky fraction \\ \hline 
       CMB-S4 & $\theta_{\rm FWHM}~[\rm arcmin]=$ & 5.5 & 2.3 & 1.4 & 1.0 & 0.7 \\ 
        & $\Delta_T~[\mu K']=$ & 21.8 & 12.4 & 2.0 & 6.9 & 
         \\ \hline
       Simons Observatory (SO) & $\theta_{\rm FWHM}~[\rm arcmin]=$ & 5.5 & 2.3 & 1.4 & 1.0 & 0.4 \\ 
        & $\Delta_T~[\mu K']=$ & 27.0 & 5.8 & 6.3 & 15.0 & 
         \\ \hline         
\end{tabular}
\caption{The CMB white noise parameters used in this analysis that match CMB-S4 and SO experimental specifications.} \label{tab:cmb_specs}
\vspace*{-0.4cm}
\end{table*}
\end{center}

\begin{center}
\begin{table}
\centering
\begin{tabular}{| c | r | c | c | c | c | c | c | c | c | c | c | c | c | c |}\hline 
       LSS Experiment & Redshift $z$ = &  0.26 & 0.38 & 0.50 & 0.64 & 0.79 & 0.96 & 1.14 & 1.35 & 1.58 & 1.84 & 2.15 \\\hline 
       LSST & ${\dd n_{\rm gal}(z)/\dd z~[\rm 1/arcmin^2]=}$ & 15.4 & 20.8 & 22.9 & 21.8 & 18.6 & 14.4 & 10.0 & 6.34 & 3.57 & 1.77 & 0.75 \\
       DESI & $\dd n_{\rm gal}(z)/\dd z~[\rm 1/arcmin^2]=$ & 0.15 &  0.18 & 0.24 & 0.36 & 0.56 & 0.47 & 0.45 & 0.36 & 0.14 & 0.02 & 0.0 \\\hline 
\end{tabular}
\caption{The galaxy number counts we use that match LSST ($m_{\rm lim}\gtrsim26$) and DESI surveys.}
\label{tab:lss_specs}
\vspace*{-0.4cm}
\end{table}
\end{center}

 \vspace*{-2cm}

\section{Detecting the moving lens effect}
\label{sec:detect}

In this section, we present strategies for the detection of the ML effect. 
We first apply a component separation method from multi-frequency CMB observations in order to minimize the contamination from frequency-dependent foregrounds. We describe the standard  ILC-cleaning method we apply in Appendix~\ref{sec:ILC-appendix}.
The ILC-cleaned maps contain all black-body signals including the primary (lensed) CMB, kSZ effect, halo lensing and the ML effect. Although reduced by around an order of magnitude from ILC-cleaning, note that our maps still contain significant residual foregrounds that dominate the CMB on sub-degree scales (see Fig.~\ref{fig:mixedmaps}).

Here we study two methods for the detection of the ML effect: pairwise transverse velocity estimation (Sec.~\ref{sec:paird}) and oriented stacking (Sec.~\ref{sec:oriented_stacking}). While aiming at the same goal and in principle containing equivalent information~\citep[in case both methods are performed optimally, see~e.g.][]{Smith:2018bpn}; in practice the two methods follow different procedures and the results can depend differently on systematic effects, feasibility on the analyses, and survey selections. Our goal is to provide 
an assessment of each method for detecting the ML effect for a realistic analysis in the presence of all contributions to the reconstructed CMB map. 

We discuss to what extent the results depend on the number of objects/pairs considered, their mass and their redshift, and what are the main obstacles for the detection implied by each strategy. 
We ultimately aim at offering guidelines for the most feasible and best possible data analysis of future experiments, and help improve current strategies for detecting and utilizing the ML signal. While the two methods follow different procedures, they use the same set of simulations, as described above.

\subsection{Detection via oriented stacking}

\begin{figure}[t!]
 \centering
\includegraphics[width=0.3\linewidth]{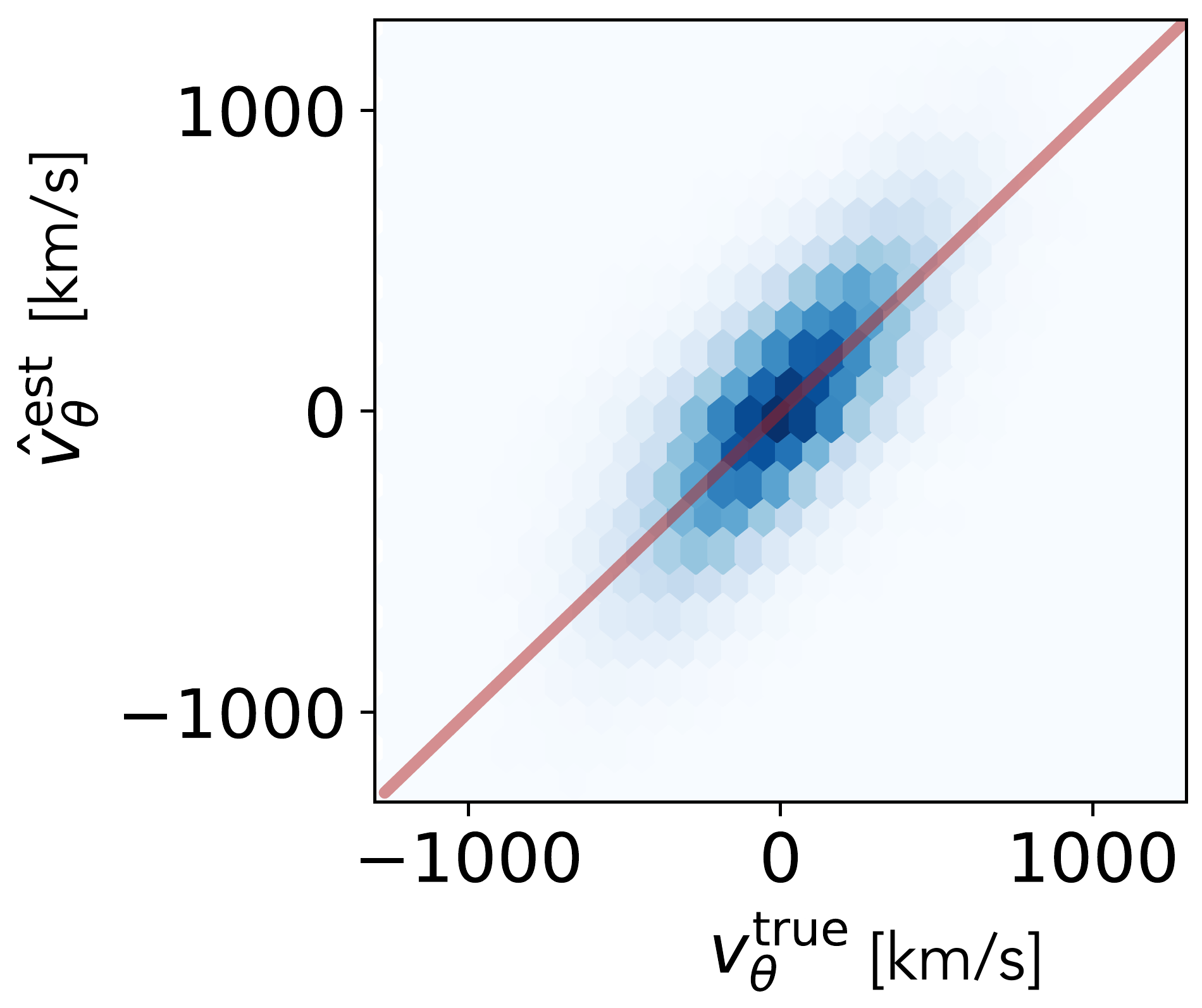}\vspace*{-0.2cm} \caption{{Comparison between reconstructed halo transverse velocities $v_\theta^{\rm est}$ with the true halo velocities from \texttt{websky} simulation $v_\theta^{\rm true}$, for one of the two transverse velocity components on 2-sphere $x=\{\phi,\,\theta\}$. The results are qualitatively identical for the  two components.} Our velocity reconstruction procedure from galaxy number count distribution is described in Sec.~\ref{sec:velocity_recon}. Estimated velocities corresponding to a LSST-like survey are shown. We find for both DESI- and LSST-like surveys similar {Pearson correlation coefficients}$^4$ of around $\sim70$ percent for transverse velocity reconstruction (around $\sim65$ percent for radial velocity reconstruction). {We also find that the velocities are over estimated for large values.}
}\label{fig:velo_recon}
 \vspace*{-0.3cm}
 \end{figure}

One of the promising ways for the first detection of the ML effect is by stacking patches of CMB maps at the locations of DM halos, after orienting each patch along the direction of the local bulk transverse velocity. While the ML signal is {typically} small to be observed per halo basis, {oriented-stacking many CMB patches  increases the prospect of detecting this signal.} The procedure involves estimating the bulk transverse-velocity field through a linear reconstruction from a galaxy survey, briefly described in Sec.~\ref{sec:velocity_recon}, followed with oriented stacking using these estimates, described in Sec.~\ref{sec:oriented_stacking}. We demonstrate the disparaging effects of different contributions to CMB on the ML detection in Sec.~\ref{sec:foreground_biases}. 

\subsubsection{Transverse velocity reconstruction from galaxy surveys}\label{sec:velocity_recon}

On linear scales, large-scale structure density and velocity fluctuations related by the continuity relation 
\be\label{eq:continuity}
\boldsymbol{\nabla}\cdot\boldsymbol{v}(\nhat)+f\,\boldsymbol{\nabla}\cdot[(\nhat\cdot\boldsymbol{v}(\nhat))\nhat]=-aHf\delta(\nhat)\,, 
\ee
where $a$ is the scale factor, $H$ is the Hubble parameter and $f=\dd \ln D(a)/\dd \ln a$ is the cosmological growth rate, $\boldsymbol{v}(\nhat)$ is the bulk 3D velocity at the line-of-sight direction $\nhat$ and $\delta(\nhat)$ is the density fluctuations. Also on these scales, fluctuations of galaxy number counts trace the density fluctuations, satisfying $\delta_g(\nhat)\simeq b_g\delta(\nhat)$, and as a result, bulk transverse velocity field could be estimated from a linear reconstruction from the galaxy density field. 

In order to estimate the galaxy velocities, we first place our galaxy catalogs on a 3D grid in real space, which yields an estimate of the 3D galaxy density field. This field can then be converted to velocities using the continuity equation in Eq.~\eqref{eq:continuity}. The second term in the left-hand side of Eq.~\eqref{eq:continuity} takes into account the linear redshift-space distortion (Kaiser effect) that affects the radial velocity component. The velocity reconstruction is hence not perfect due to redshift distortions, as well as finite number count of galaxies (i.e. shot noise), non-linearity of the galaxy over-density and the finite volume observed. We find our reconstructed velocities to be $\sim65-70$ percent correlated with the true halo velocities for LSST- and DESI-like galaxy survey specifications;\footnote{Here, we use the Pearson correlation coefficient defined as 
\be
r=\frac{\sum_i (\hat{v}^{\rm est}_{x,i}-\bar{\hat{v}}^{\rm est}_x)(v^{\rm true}_{x,i}-\bar{v}^{\rm true}_{x})}{\sqrt{(\sum_i (\hat{v}^{\rm est}_{x,i}-\bar{\hat{v}}^{\rm est}_x)^2\sum_i(v^{\rm true}_{x,i}-\bar{v}^{\rm true}_{x})^2}}
\ee 
where $i$ runs over halos, $x\in\{\theta,\phi\}$, and $\hat{v}^{\rm est}_{x,i}$ (${v}^{\rm true}_{x,i}$) correspond to estimated (true) velocities.} in agreement with the fidelity of the velocity reconstruction achieved in earlier studies~\citep[e.g.][]{ACTPol:2015teu}. We show a comparison between reconstructed halo velocities from a LSST-like survey specifications and true halo velocities from \texttt{websky} simulation in Fig.~\ref{fig:velo_recon}.

\vspace*{1cm}
\subsubsection{Oriented stacking}\label{sec:oriented_stacking}

We stack grids of CMB patches around the line-of-sight direction of halos in the halo catalog after rotating them to align the estimated velocities. We map the aligned patches onto a $N_p\times N_p$ grid of pixels whose centers evenly cover the range $[-\lambda r/r_s,,+\lambda r/r_s]$ in two orthogonal directions on the 2-sphere. Our algorithm is described in Appendix~\ref{sec:stacking-appendix}. Once rotated with Eq.~\eqref{eq:stackingrot}, we stack each of these $N_p\times N_p$ pixels individually over the catalog. We choose $N_p=21$ and $\lambda=2.5$ unless otherwise stated, which sufficiently captures the characteristic profile of the ML signal as can be seen from the leftmost panel in Fig.~\ref{fig:stacked_profiles1}, for example. We avoid patches larger than than $\lambda\sim2.5$, for which we have found the contribution to the stacked profiles remain dominated by foregrounds \citep{Hotinli:2023ywh}.

\subsubsection{Foreground biases}\label{sec:foreground_biases}

\begin{figure}[t!]
\label{fig:ml_recon}
 \centering
\hspace*{-1cm}
\includegraphics[width=1.1\linewidth]{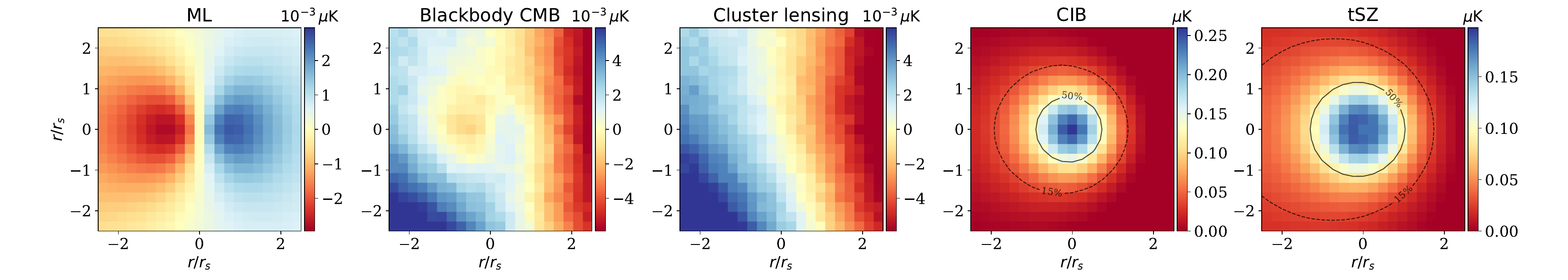}
\vspace*{-0.5cm}
\caption{Stacked CMB patches for a CMB-S4-like resolution and noise. The Number of CMB patches, equal in each stack, matches the halo number count anticipated from an LSST-like galaxy survey.  The color bars correspond to the mean temperature at each pixel after averaging. The leftmost panel corresponds to the moving lens signal in isolation. The second panel to the left corresponds to the total black-body signal in the CMB {map}, including kSZ, {halo} lensing as well as the primary lensed CMB. The moving lens is apparent from the second stack although {halo} lensing contributes a significant gradient largely obscuring the dipolar ML signature. The middle panel shows the {halo} lensing signal in isolation {(note that here, the direction of the gradient is accidental and not correlated with the velocity)}. The right two panels correspond to the contribution of CIB and tSZ effects to the ILC-cleaned CMB maps {we describe in Sec.~\ref{sec:foreground_biases}}, which are still orders of magnitude large than the ML signal. The contour lines on this panels correspond to pixels corresponding to 50~and~15 percent of the central pixel value, demonstrating the anisotropic profile of the residual foregrounds aligned with the transverse velocity on the stacked patches.} \label{fig:stacked_profiles1}
\end{figure}

Figs.~\ref{fig:stacked_profiles1}-\ref{fig:SNRMz} demonstrate the results from our stacking analysis. We find that foregrounds such as halo lensing, tSZ and CIB, introduce significant gradients on the stacked patches aligned with the transverse velocity direction. The three panels from right in Fig.~\ref{fig:stacked_profiles1} show contributions to stacks from halo lensing, CIB and tSZ respectively. {The CIB and tSZ contributions are calculated by taking the difference between ILC-cleaned CMB maps including all foregrounds, and similarly--ILC-cleaned CMB maps however omitting either CIB or tSZ foregrounds, respectively. We generate the map of halo lensing by painting the lensing signal from each halo using \texttt{AstroPaint}. We define the halo lensing signal as $\Theta_{\rm HL}^{(h)}(\nhat)=\boldsymbol{\nabla}T^u(\nhat)\cdot\boldsymbol{\nabla}\phi(\nhat)$ where $\phi=-(2/c^2)\int_0^{\chi_*} \dd\chi[(\chi-\chi_*)/ (\chi\chi_*)] \Psi_h(\chi\nhat)$, where $\chi_*$ is the comoving distance to recombination surface and $T^u(\nhat)$ is the \textit{unlensed} CMB temperature from \texttt{websky}.} We find halo lensing corresponds to the dominant contribution from black-body CMB, leading to an overall factor $\sim2$ larger gradient compared to ML signal.
The residual contributions to the stacks from CIB and tSZ effects after ILC-cleaning also lead to contributions that are multiple orders of magnitude larger than {the} ML and halo lensing, with a similar  gradient that is a factor $\sim5$ larger than ML signature. 

The significance of these contributions (or biases) we note here \citep[and also in][]{Hotinli:2023ywh} suggest that the foreground contributions to the CMB needs to be mitigated more effectively (for example by removing the frequency dependent CIB and tSZ foregrounds with a more effective cleaning method) or be modelled, where model parameters must be marginalized. In the following section we demonstrate the prospects of detecting the ML effect if the latter is achieved. We leave studying the improvements of more advanced ILC-cleaning methods such as discussed in~\citep[e.g.][]{McCarthy:2023hpa,McCarthy:2023cwg,2009A&A...493..835D,2009ApJ...694..222C,2011MNRAS.410.2481R} to upcoming work, although see Appendix~\ref{app:deproj} for a preliminary analysis. {That brings a marginal ($10\%$) improvement to these results.}

\subsubsection{Forecasts and bias modelling}\label{sec:forecasts_fisher}
In this section, we use the simulated ILC-cleaned CMB map{s} to assess the prospects of detecting the ML signal in the presence of detrimental contributions to stacks from foregrounds shown above.
We adopt an analytical approach to investigate which foreground or competing signal has the largest impact on the detection. {We find CIB and tSZ foregrounds correspond to the dominant contribution to reducing the prospects of an unambiguous detection of the ML signal.} 

The signal we consider, $\boldsymbol{\Delta S}$, corresponds to the difference between patches produced by orienting the stacks {of ILC-cleaned CMB maps} based on the estimated transverse velocities, and patches produced by stacking the same images with random orientation. {In order to assess the contribution from ML signal, primary CMB and different foregrounds to this statistic, we approximate} $\boldsymbol{\Delta S}$ as: 
\be\label{eq:signal_templates}
{\boldsymbol{\Delta \tilde{S}}}=A_{\rm ML}\boldsymbol{\Delta S}_{\rm ML}+A_{\rm lCMB}\boldsymbol{\Delta S}_{\rm lCMB}+A_{\rm kSZ}\boldsymbol{\Delta S}_{\rm kSZ}+A_{\rm HL}\boldsymbol{\Delta S}_{\rm HL}+A_{\rm CIB}\boldsymbol{\Delta S}_{\rm CIB}+A_{\rm tSZ}\boldsymbol{\Delta S}_{\rm tSZ}\,,
\ee 
{where $\boldsymbol{\Delta \tilde{S}}$ is a template we construct from summing the contributions from different components to $\boldsymbol{\Delta {S}}$.} Here $\boldsymbol{\pi}=\{A_{\rm ML},A_{\rm lCMB},A_{\rm kSZ},A_{\rm HL},A_{\rm CIB},A_{\rm tSZ}\}$ is a set of parameters that correspond to amplitudes of various contributions to $\Delta \boldsymbol{S}$ which we calculate from the \texttt{websky} simulations. These are the ML effect, lensed primary CMB, kSZ effect, halo lensing, CIB, and tSZ effect, respectively. {For components other than ML, we calculate these contributions by taking the difference between the total stack $\Delta \boldsymbol{S}$ obtained from ILC-cleaned CMB maps, and ones that follow the same procedure except exclude each component, as described in Sec.~\ref{sec:foreground_biases} CIB and tSZ. While $\Delta \boldsymbol{\tilde{S}}$ is not strictly equal to $\boldsymbol{\Delta S}$, since the performance of ILC-cleaning depends on the combination of all foregrounds in the observed map, this procedure allows us to produce a tentative template  for each contribution and assess the degeneracy between different components and the ML signal via a Fisher matrix analysis, which we describe below.} We find experimental noise and point sources mainly contribute as sources of noise, hence do not bias the ML detection.

\begin{figure}[t!]
 \centering
\includegraphics[width=0.3\linewidth]{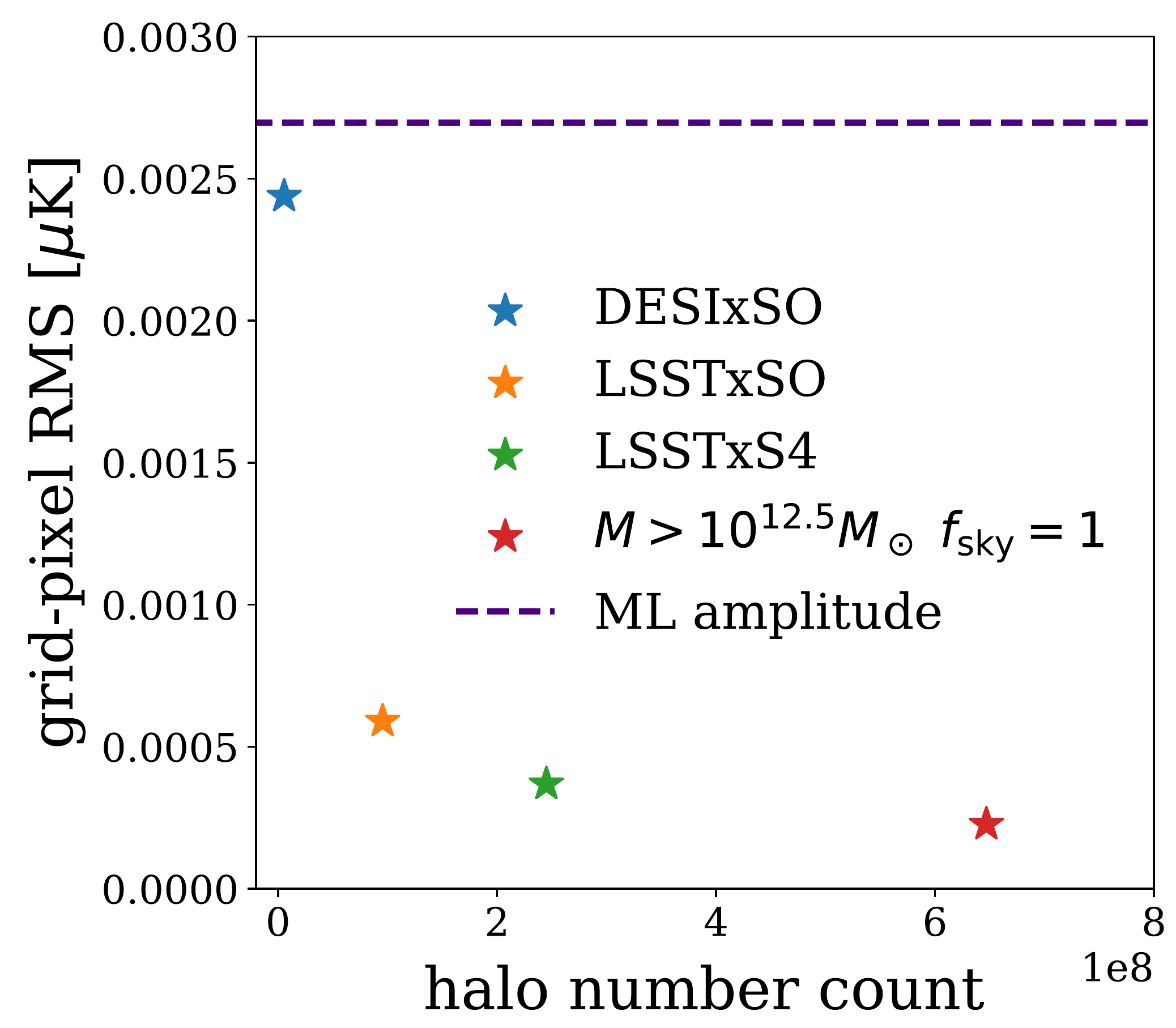}
\vspace*{-0.2cm}
\caption{The mean RMS of pixels in stacks of ILC-cleaned CMB matching CMB-S4 and SO experimental specifications, for varying halo number count matching DESI and LSST surveys. Star markers correspond to various experiment combinations with varying halo number count and CMB specifications. The RMS is dominantly sourced by the lensed primary CMB and vary below the order of perfect of the mean throughout the pixels in the grids. The RMS values can be seen to be lower than the ML signal amplitude (here parametrised by the maximum value of the ML profile in $\mu K$ with the purple dashed line) for all experiments we consider. }
\label{fig:stacked_profiles2}
\end{figure}

{We calculate the noise for each pixel via the `delete-d' jack-knife method \citep{Escoffier:2016qnf} which we perform by rerunning our analysis a second time after calculating each stack, this time recording the difference between each pixel of the patches and the total stack, in quadrature. We quote the mean of this difference as our pixel variance.} The pixel variance is shown in Fig.~\ref{fig:stacked_profiles2} for a range of CMB and LSS experiments. {As evident from this figure, we find halo number counts play a crucial role in improving the pixel variance for the stacking analysis, especially when going from DESI-like halo number counts to those expected from LSST.}

We define the signal-to-noise (SNR) for detecting the ML signal as the error on the amplitude $A_{\rm ML}$, after marginalizing over the remaining amplitudes. To this aim, we define an ensemble-information matrix as
\be\label{eq:fisher}
\mathcal{F}_{ik}=\sum\limits_{\alpha}\! \left[\frac{\partial \boldsymbol{\Delta S}_{\alpha}}{\partial \pi_i}(\boldsymbol{\Delta S}\! +\!\boldsymbol{N})_{\alpha}^{-1}\frac{\partial \boldsymbol{\Delta S}_{\alpha}}{\partial \pi_k}(\boldsymbol{\Delta S}\!
+\!\boldsymbol{N})_{\alpha}^{-1}\right]\,,
\ee
where signal and noise vectors satisfy
\be
{\boldsymbol{\Delta S}}=\{\Delta T_{0},\Delta T_{1}, \ldots, \Delta T_{N_p\times N_p}\}\,\,\,\,\,\,\,{\rm and}\,\,\,\,\,\,\,{\boldsymbol{N}}=\{\sigma(\Delta T_{0}),\sigma(\Delta T_{1}), \ldots, \sigma(\Delta T_{N_p\times N_p})\}\,,
\ee
and the noise vector $\boldsymbol{N}$ corresponds to the errors on each pixel of the ILC-cleaned CMB patches including all foregrounds (after removing randomly oriented patches). The subscript indices span the $N_p\times N_p$ grid points of the rotated CMB patches we described earlier, and we take $N_p=21$. {Here $\boldsymbol{\pi}$ array consists of the set of parameters, as defined above.}

\begin{figure}[t!]
\centering
\includegraphics[width=0.9\linewidth]{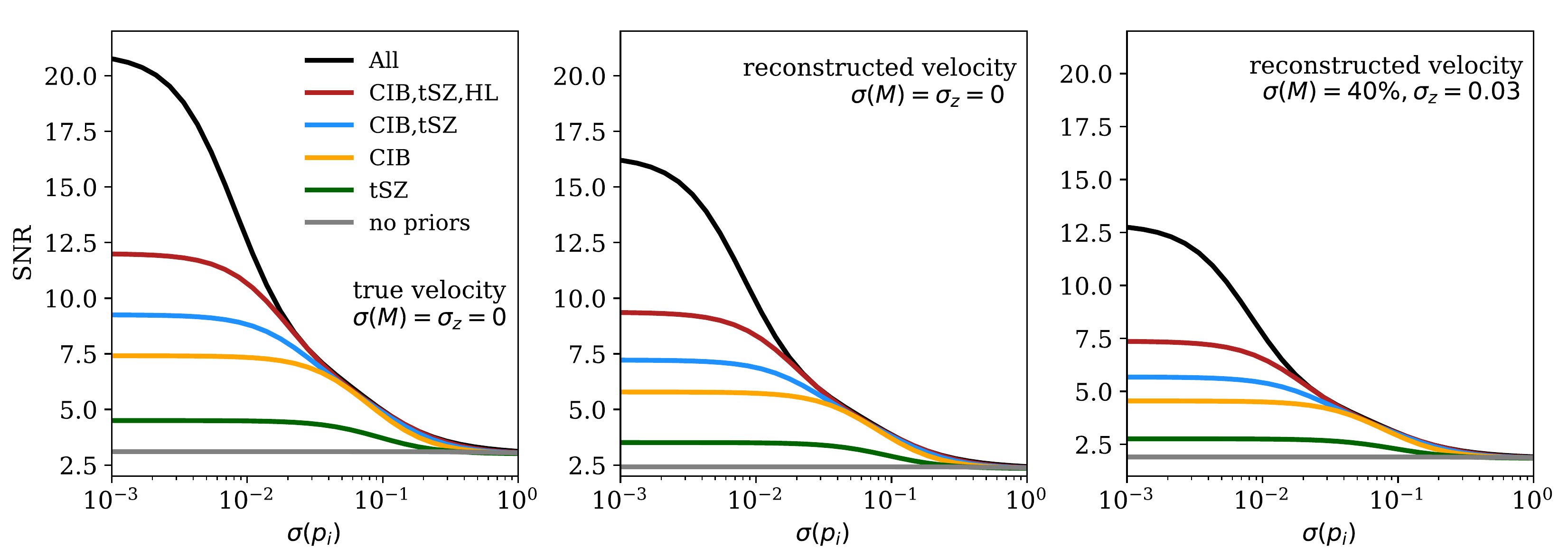}
\vspace*{-0.33cm}
\caption{The anticipated signal-to-noise ratio (SNR) for the detection of the ML effect (which we define as the SNR on the $A_{\rm ML}$ parameter defined in the text), for CMB-S4 and LSST. The $x$-axes on these plots correspond to priors on the amplitudes of most significant CMB foregrounds: $\{A_{\rm CIB},\,A_{\rm tSZ},\,A_{\rm HL}\}$, for CIB, tSZ and halo lensing, respectively.
The degrading effect of the degeneracy between the ML signature and the other CMB signals can be seen from the right end of each panel: in the absence of prior knowledge of the anticipated biases, the SNR of detecting the ML effect is small. The black curves correspond to assuming priors on all of the other CMB signals or foregrounds defined in Eq.~\eqref{eq:signal_templates}, and the SNR can be seen to improve as stronger priors are set. The green and orange curves correspond to assuming priors on the tSZ and CIB biases alone, which can be seen as the dominant contributions to the reducing the ML detection. The blue curves are from assuming (equal percent) priors on both CIB and tSZ biases. The red curve corresponds to similarly assuming priors on $\{A_{\rm CIB},\,A_{\rm tSZ},\,A_{\rm HL}\}$. We omit showing results for kSZ, lensed primary CMB and halo lensing alone, as these are within few percents of the gray curves, which assume no priors.}\label{fig:SNR_from_marginalization}
\vspace*{-0.1cm}
 \end{figure}
 
 \begin{figure}[b!]
 \vspace*{0.5cm}
\includegraphics[width=\linewidth]{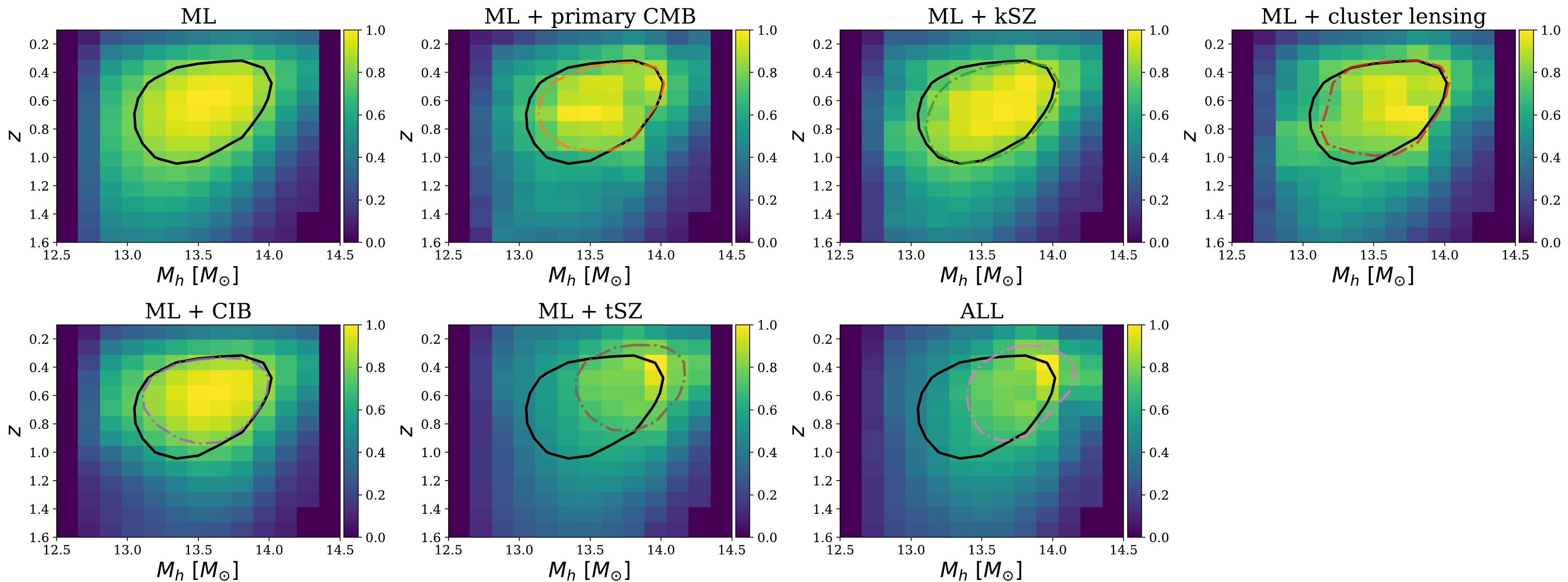}\caption{The fractional SNR contribution from halos within bins of redshift and mass to the total detection SNR with stacking, {for CMB-S4-- and LSST--like experiments.} Different panels correspond to considering {the} ML signal in isolation (top left panel) and together with {one} other CMB signal (the right three panels in the top and left two panels in the bottom row). The right-most bottom panel corresponds to ILC-cleaned CMB maps including all foregrounds as well as the ML effect. 
The contour lines correspond to the area that includes 75$\%$ of the total SNR. The black solid {(colored dot-dashed)} contour lines correspond to taking the same percentile in SNR, considering ML signal in isolation {(considering the combination of signals labelled at the top of each panel)}. The contribution to the total SNR can be shown to shift towards lower redshift higher mass halos with the inclusion of other CMB foregrounds. Masses are in logarithmic units in all plots.}
\label{fig:SNRMz}
\end{figure}

Fig.~\ref{fig:SNR_from_marginalization} demonstrates the anticipated SNR of the detection of the ML effect (i.e. defined as the SNR on the $A_{\rm ML}$ parameter). The $x$-axes on these plots correspond to enforced priors on the amplitudes of the remaining CMB foregrounds. The degrading effect of the degeneracies between the ML signature and the other CMB foregrounds can be seen from the right end of each panel. In the absence of prior knowledge of the anticipated biases, the SNR of detecting the ML effect is small {($\simeq 3$)}. 
It is therefore mandatory to find a better cleaning method to be able to use the stacking analysis to recover peculiar velocities. {What will a better method need to address?} Fig.~\ref{fig:SNR_from_marginalization} shows that the main limiting factor in the oriented stacking analysis  is the residual  CIB, followed by the tSZ, since these signals have the same directionality than the moving lens (though for different physical reasons) {as shown with the rightmost two panels in Fig.~\ref{fig:ml_recon}.
The use of the reconstructed velocity instead of the true halo velocity reduces the maximum attainable S/N from 20 to 15, and the  uncertainty on mass estimate and redshifts further reduces the maximum SNR to $\sim10$. {Indeed, better strategies for velocity reconstruction are already available~\citep[e.g.][and references therein]{Hadzhiyska:2023nig,Guachalla:2023lbx} including via non-linear methods and efforts to improve mass reconstruction are already ongoing~\citep[e.g.][]{Bayer:2022vid}. 

From our maps, we can also assess which types of objects mainly impact the SNR. Indeed, larger objects at low redshifts produce the largest moving lens signal. However, there are fewer of them, and they also have larger foreground signals. 
So we analyzed which objects contribute the most, given the current strategy for map cleaning. 
Results are shown in Fig.~\ref{fig:SNRMz}. 

The contribution to the total SNR can be shown to shift towards lower-redshift and higher-mass halos with the inclusion of other CMB foregrounds. The effect is most pronounced from adding the tSZ signal.
Our analysis so far suggests the bulk of the SNR will be provided from $M_h\gtrsim3\!\times\!10^{13}M_\odot$ halos due to detrimental effects from foregrounds, with lower redshift objects being increasingly important.
Considerations of potential survey strategies and alternative foreground cleaning methods can change these results.
We will explore such possibilities in future works. 

\vspace*{-0.2cm}
\subsection{Detecting the pairwise velocity}

Next, we consider pairwise transverse-velocity detection from the ML effect. As discussed above, the method of pairwise-velocity detection has been valuable for the detection of kSZ effect in the past years~\citep[see~e.g.][]{Hall:2014wna,Schaan:2016ois}. One of the benefits of pairwise-velocity detection is that the method does not rely on the estimation of the halo bulk velocities from a galaxy survey. Since due to gravity any two halos are more likely to be moving towards each other, this method naturally allows reconstructing the non-zero mean pairwise velocity signal from pairs of halos at $\mathcal{O}(10-100)$~Mpc distances,  given sufficient statistical power. 
By virtue of the fact that the relevant peculiar velocity here is not the total peculiar velocity but rather the projection of it in the direction of the vector connecting the pair, we expect the  CIB and tSZ bias to be reduced or eliminated. In what follows, we test our speculation and assess the expected SNR for the detection.

In order to measure the pairwise velocities for a given combination of CMB experiment and galaxy catalog, we adopt the estimator presented in \citep{Yasini:2018rrl}. We first estimate the halos' individual peculiar  transverse velocities from the same CMB maps we use for the stacking analysis. In Sec.~\ref{sec:indv} we describe how we recover individual transverse velocities from CMB patches. In Sec.~\ref{sec:paird} we then use these individually-determined velocities to infer the estimated pairwise transverse-velocity signal. We present our results for different experimental configurations, halo mass and redshift cuts in Sec.~\ref{sec:results}. {Note that similar to oriented stacking, pairwise velocity estimation is also computationally challenging due to large, $\sim\mathcal{O}(10^8)$, number of halo pairs needed for the analysis to be representative of CMB-S4 and LSST, for example. We have made our parallelized and efficient code available at \hyperlink{https://github.com/selimhotinli/moving_lens}{selimhotinli/moving\_lens}.

\subsubsection{Detecting individual velocities}
\label{sec:indv}

Here we apply a matched filter to patches of CMB to estimate transverse velocity components $v_{\phi,i}(\nhat)$ and $v_{\theta,i}(\nhat)$. The matched filter we use is a simple modification of the one presented in~\cite{Hotinli:2019wdp}, and has the Fourier-space form 
\be
\Psi^\theta_i(\boldsymbol{\ell})=N_{i}\,\frac{|\hat{\theta} \cdot\boldsymbol{\beta}_i (\boldsymbol{\ell})|^2}{C_\ell^{\Theta\Theta,\rm obt}}\,,
\ee 
(and similarly for $\hat{\phi}$). Here, the `obt' superscript indicates the obtained CMB variance including residual foregrounds and noise after ILC cleaning, and $N_i=\int\dd^2\boldsymbol{\ell}|\Psi^\theta_i(\boldsymbol{\ell})^2|C_\ell^{\Theta\Theta,\rm obt}$ is the normalisation of the filter. The velocity estimator can then be calculated as done in~\cite{Hotinli:2019wdp}, and takes the form 
\be
\tilde{v}^\theta_i(\boldsymbol{x})=\frac{\int\dd^2 \boldsymbol{x}\, \Psi^\theta_i(\boldsymbol{x})\tilde{\Theta}^{\theta}(\boldsymbol{x})}{\int\dd^2\boldsymbol{x}|\Psi^\theta_i(\boldsymbol{x})|^2}\,,
\ee 
in flat-sky coordinates $\boldsymbol{x}$, where $\tilde{\Theta}^{\theta}(\boldsymbol{x})$ is the obtained CMB patch \textit{filtered} to satisfy 
\be
\tilde{\Theta}^{\theta} (\boldsymbol{\ell}) =N_i\,\frac{\hat{\theta} \cdot\boldsymbol{\beta}_i (\boldsymbol{\ell})}{ C_\ell^{\Theta\Theta,\rm obt}}\,\Theta^{\rm obt}(\boldsymbol{\ell})\,,
\ee 
in Fourier space.
The estimator for $\hat{v}_{\phi,i}(\boldsymbol{x})$ can be obtained similarly by trading $\hat{\phi}$ with $\hat{\theta}$ in the above expressions. Here, we apply our matched filter to patches of CMB centered at \texttt{websky} halo centers and obtain estimations for the two velocities $\hat{v}_{\phi,i}$ and $\hat{v}_{\theta,i}$. Next, we apply our pairwise-velocity estimator to pairs of halos to measure the velocity signal.

\begin{figure*}[t!]
\centering
\includegraphics[width=1\linewidth]{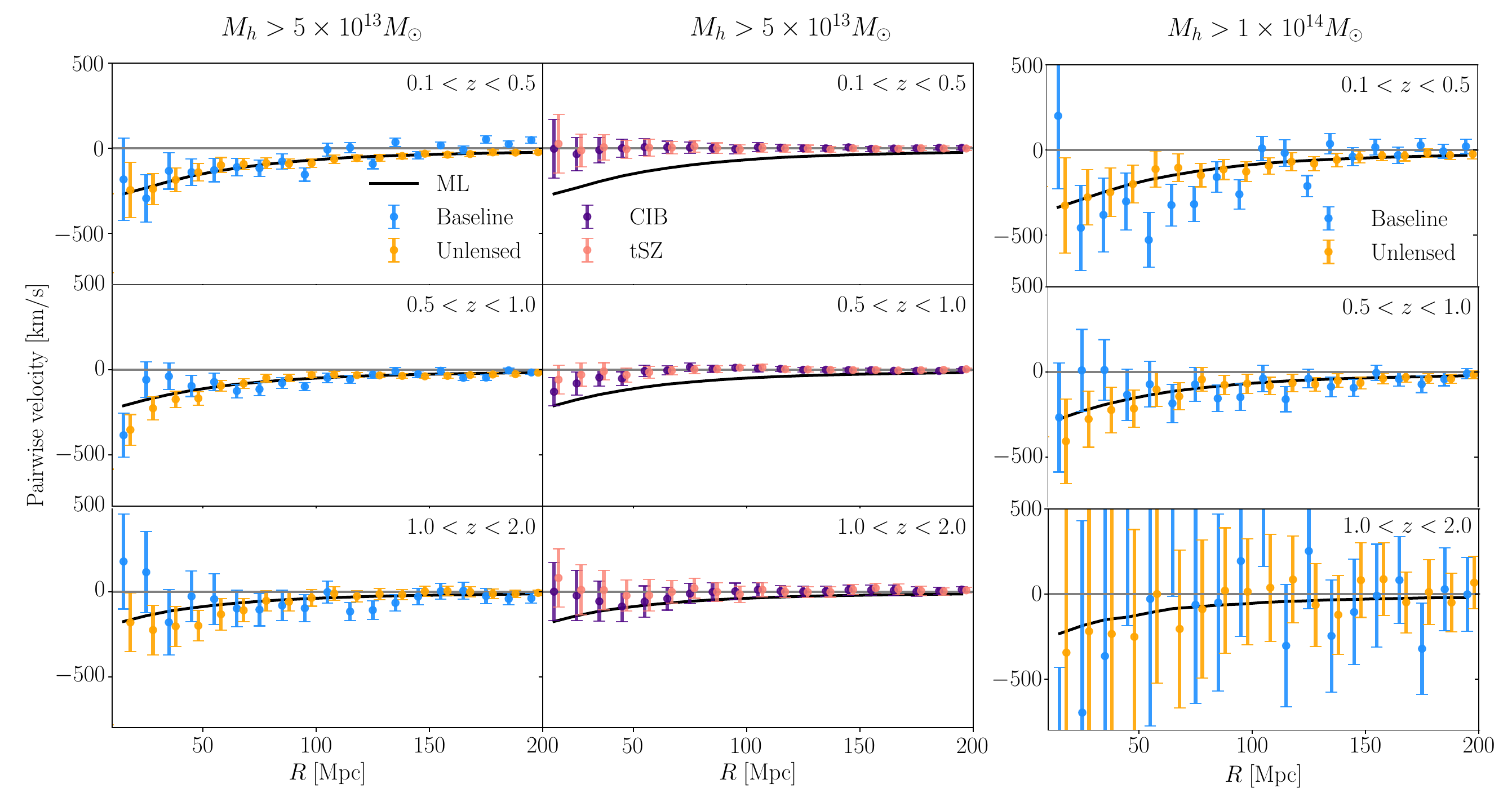}
\vspace*{-0.7cm}
\caption{Pairwise transverse-velocity signal and measurement error at different redshifts for the combination of CMB-S4 and LSST surveys.  The black lines on all panels correspond to the pairwise transverse-velocity signal from maps including simulations of ML effect alone, which we verifies to follow the theoretical prediction for the pairwise velocity signal. The top panels corresponds to results of pairwise-velocity estimation from low-redshift halos at $z\in[0.1-0.5]$. The middle and lower panels corresponds to redshift bins $z\in[0.5,1.0]$ and $z\in[1.0,2.0]$, respectively. The blue data points on the left and right panels correspond to results from pairwise-velocity estimation using our \baseline~ILC-cleaned CMB which includes all foregrounds, with white noise matching that of CMB-S4. The orange data points on the same panels correspond to an identical analysis except omitting the contribution to the {observed} CMB maps from cluster lensing. The left (right) panels correspond to applying a halo mass cut of $M_h>5\times10^{13}M_\odot$ ($M_h>1\times10^{14}M_\odot$). The purple and pink data points on the middle panels correspond to contributions to the pairwise-velocity signal from frequency-dependent CIB and tSZ foregrounds, respectively, for $M_h>5\times10^{13}M_\odot$. We calculate these purple and pink data points by performing ILC-cleaning with identical foregrounds except omitting CIB and tSZ, respectively, and subtracting these maps from our \baseline~ILC-cleaned maps, which include all foregrounds. The remaining maps leading to purple and pink data points capture the contribution to the \baseline~ILC-cleaned maps from CIB and tSZ foregrounds in isolation, respectively. We calculate the error bars with the jackknife method, as described in the text. The total number count of halos satisfy the corresponding mass cuts, as well as the sky fraction, which match the anticipated values from joint analysis of CMB-S4 and LSST surveys. For $M_h>5\times10^{13}M_\odot$, we find the total detection signal-to-noise ($\mathcal{S}/\mathcal{N}$) of the pairwise velocity to be 9.4 (14.1) for our \baseline~(\unlensed) results, which we detail in Table~\ref{tab:SNR_1}. For $M_h>1\times10^{14}M_\odot$, we find corresponding $\mathcal{S}/\mathcal{N}$ satisfy 6.3 (7.9), for baseline (unlensed), which we show in Table~\ref{tab:SNR_2}.}\label{fig:velo_estimate_pw_1}
\end{figure*}

\vspace*{-0.5cm}
\subsubsection{Estimating the pairwise velocities}
\label{sec:paird}

Following~\cite{Yasini:2018rrl}, the mean pairwise velocity between all pairs $\{i,j\}$ at distance $r$ from each other can be defined as 
\be
v_{\langle ij\rangle}(r)\equiv\langle\boldsymbol{v}_{ij}(r)\cdot\hat{\boldsymbol{r}}_{ij}\rangle
\ee
where $\boldsymbol{v}_{ij}=\boldsymbol{v}_i-\boldsymbol{v}_j$ is the relative 3-dimensional velocity between two halos and $\boldsymbol{r}_{ij}=r\,\hat{\boldsymbol{r}}_{ij}=\boldsymbol{r}_j-\boldsymbol{r}_i$ is the distance between a pair of halos at locations $\boldsymbol{r}_i$ and $\boldsymbol{r}_j$. The estimator for $v_{\langle ij\rangle}$ can be found to satisfy
\be\label{eq:pairwise_est}
\tilde{v}_{\langle ij\rangle}=\frac{\sum_{ij}(\tilde{\boldsymbol{t}}_i-\tilde{\boldsymbol{t}}_j)\cdot\boldsymbol{q}_{ij}}{\sum\limits_{ij}|\boldsymbol{q}_{ij}|^2}\,, 
\ee
where $\tilde{\boldsymbol{t}}_i=\hat{\boldsymbol{r}}_i\times(\tilde{\boldsymbol{v}}_i\times\hat{\boldsymbol{r}}_i)$ and $\tilde{\boldsymbol{t}}_j=\hat{\boldsymbol{r}}_j\times(\tilde{\boldsymbol{v}}_j\times\hat{\boldsymbol{r}}_j)$ are transverse velocities and $\boldsymbol{q}_{ij}\equiv[2\hat{\boldsymbol{r}}_{ij}-\hat{\boldsymbol{r}}_{i}(\hat{\boldsymbol{r}}_{ij}\cdot\hat{\boldsymbol{r}}_i)-\hat{\boldsymbol{r}}_{j}(\hat{\boldsymbol{r}}_{ij}\cdot\hat{\boldsymbol{r}}_j)]$. Here, we have used tilde on estimated observables. 

We use Eq.~\eqref{eq:pairwise_est} with the estimated individual velocities as described in Sec.~\ref{sec:indv}. We calculate the covariance of estimated velocities $\tilde{v}_{\langle ij\rangle}$ via the `delete-d' jacknife method~\citep{Escoffier:2016qnf}. We divide our full sample of velocities into a 100 sub-samples. We select one of these sub-samples and delete it from our full sample. We calculate mean pairwise velocity given redshift bin for the 99 out of 100 sub-samples. We repeat this process for each of the 100 sub-samples and calculate the resulting variance. Next, we assess the fidelity of the pairwise transfer-velocity measurement for a range halo masses, redshifts and experimental considerations.

\subsubsection{Forecasts}\label{sec:results}

Fig.~\ref{fig:velo_estimate_pw_1} corresponds to our forecasts for an analysis of CMB-S4--like survey joint with a LSS survey matching specifications of LSST. The blue data points throughout correspond to our \baseline~choice of CMB maps following standard ILC-cleaning taking into account frequency dependent CIB and tSZ foregrounds, white noise matching CMB-S4, point sources, and frequency-independent effects including halo lensing, kSZ, in addition to primary (lensed) CMB, as we have done for oriented stacking analysis above. The orange data points in both figures correspond to following up an identical analysis, however not taking into account the halo lensing foreground.

{The panels on the middle column in Fig.~\ref{fig:velo_estimate_pw_1}} show the contributions to the measured pairwise velocities from CIB or tSZ foregrounds after ILC cleaning, which we calculate by repeating our baseline analysis while omitting CIB or tSZ foreground entirely; and subtracting the resulting velocity estimates from our baseline velocity estimates, shown with purple and pink data points. For the mass and redshift ranges we consider, we find CIB and tSZ foregrounds do not induce a significant bias to the transverse-velocity reconstruction, however they provide a significant contribution to the estimator variance. 

In Fig.~\ref{fig:velo_estimate_pw_1} we show results for a halo mass cut satisfying $M_h>5\times10^{13}M_\odot$ and $M_h>1\times10^{14}M_\odot$. The three rows of panels correspond to lower $z\in[0.1,0.5]$, middle $z\in[0.5,1.0]$, and higher $z\in[1.0,2.0]$ redshift ranges from top to bottom. For the specifications of CMB-S4 and LSST joint analysis, we find the lower redshift bin with $M_h>5\times10^{13}M_\odot$ contains $\sim1$M halos with $\sim450$M pairs within comoving distance $R<200$Mpc. Our middle (higher) redshift window contains $\sim1.5$M ($\sim1$M) halos with $\sim400$M ($\sim70$M) pairs satisfying $R<200$Mpc. For a higher halo mass cut satisfying $M_h>10^{14}M_\odot$, we find the lower redshift bin contains $\sim350$K halos with $\sim50$M pairs, middle redshift bin contains $\sim500$K halos with $\sim30$M pairs and high redshift bin contains $\sim200$K halos with $\sim2.5$M pairs satisfying $R<200$Mpc.

In order to assess the detection significance of pairwise velocities we define a chi-square statistic as 
\be\label{eq:chi-square}
\chi^2=(\tilde{v}_{\langle ij\rangle}-v_{\langle ij\rangle}^{\rm model})^t\mathcal{C}^{-1}(\tilde{v}_{\langle ij\rangle}-v_{\langle ij\rangle}^{\rm model})
\ee
where $\tilde{v}_{\langle ij\rangle}$ is the estimated pairwise transverse velocity (for a given $R$ bin), $v^{\rm model}_{\langle ij\rangle}$ is the model prediction of the signal and $\mathcal{C}$ is the covariance of the estimated pairwise velocity signal, calculated using `d-delete' jack-knife method detailed in Sec.~\ref{sec:paird}. In our analysis we calculate three chi-square statistic setting $v_{\langle ij \rangle}^{\rm model}$ one of $(v_{\langle ij\rangle}^{\rm bf},0,\tilde{v}^{\rm noML}_{\langle ij\rangle})$. Here $v_{\langle ij\rangle}^{\rm bf}$ corresponds to our `best fit' simulation prediction of the true pairwise transverse-velocity signal from repeating our analyses including only the ML signal, and `0' represents the true null condition in the absence of any pairwise-velocity signal. The term $\tilde{v}^{\rm noML}_{\langle ij\rangle}$ corresponds to the estimated pairwise velocities in the absence of ML effect. {Note that Eq.~\ref{eq:chi-square} includes the bin-to-bin correlation errors that are not displayed in Fig.~\ref{fig:velo_estimate_pw_1}}.

We define the detection SNR as
\be
{\rm SNR}=\sqrt{\Delta\chi^2}=\sqrt{\chi_{\rm null}^2-\chi_{\rm bf}^2}\,,
\ee
where we set $v_{\langle ij\rangle}^{\rm model}=0$ ($=v_{\langle ij\rangle}^{\rm bf}$) when calculating $\chi^2_{\rm null}$ ($\chi_{\rm bf}^2$). This statistic assesses whether the estimated pairwise velocities can be better modelled by the true pairwise-velocity signal rather than zero velocity. Since our results depend significantly on the foregrounds, we also define a separate statistic,
\be
\Delta \chi^2_{\rm noML}=\chi_{\rm null}^2-\chi_{\rm noML}^2\,,
\ee
where we set $\tilde{v}^{\rm model}_{\langle ij\rangle}$ equal to $\tilde{v}^{\rm noML}_{\langle ij\rangle}$. For cases $\Delta \chi^2_{\rm noML}>0$, this may indicate significant contribution to pairwise-velocity estimates from residual foregrounds other than the ML effect; which can in principle be a better fit to the data than the null condition. Particularly in case $({\rm SNR})^2\sim{\Delta \chi^2_{\rm noML}}$ (or $<{\Delta \chi^2_{\rm noML}}$), the residual foreground contribution to the pairwise-velocity estimate is a similarly good (or better) fit to data than the underlying pairwise-velocity signal. In what follows we omit assigning ${\rm SNR}$ to data in case $\tilde{v}^{\rm model}_{\langle ij\rangle}$ to avoid interpreting velocity estimate residuals (or biases) as signal.   

Our signal-to-noise (${\rm SNR}$) results for a CMB-S4 and LSST--like joint analysis are shown in Tables~\ref{tab:SNR_1}~and~\ref{tab:SNR_2} for halo mass cuts $M_h>5\times10^{13} M_\odot$ and $>10^{14}M_\odot$ {respectively}. For both cases we find the halo lensing to be a significant source of confusion and a potential bias to the pairwise velocity estimates from the ML effect. Furthermore, as can also be seen from Fig.~\ref{fig:velo_estimate_pw_1}, we find the advert effect of halo lensing becomes more pronounced with increasing redshifts, as our \baseline~results deviate from the transverse-velocity signal (black solid lines) more significantly for lower panels. Removing halo lensing from CMB maps, however, leads to a significant improvement for velocity estimates as can be seen form the data points labelled \unlensed. As a result, we find for our baseline assumptions, data points from pairs with higher comoving distance separations give ${\Delta \chi^2_{\rm noML}}>0$, which we exclude from our total signal-to-noise results. 

\begin{table*}
\centering
\begin{tabular}{| r | c | c | c | c | c | c | c | c | c | c |}\hline
    SNR & 
       \multicolumn{3}{c}{$z\in[0.1,0.5]$} & 
       \multicolumn{3}{|c|}{$z\in[0.5,1.0]$} & 
       \multicolumn{3}{c|}{$z\in[1.0,2.0]$}  & \\
       $R\in$ & $<150{\rm Mpc}$ & $>150{\rm Mpc}$ & All & $<150{\rm Mpc}$ & $>150{\rm Mpc}$ & All & $<150{\rm Mpc}$ & $>150{\rm Mpc}$ & All & Total \\ \hline 
        \baseline~& 5.17 & - & 5.17 &  6.31 & 2.84 & 7.11 & 3.14 & 0.99 & 3.25 & 9.4\\
        \unlensed~& 9.2 & 3.82 & 9.72 & 8.38 & 3.76 & 9.23 &  4.11 & 1.71 & 4.38 & 14.1 \\ \hline
\end{tabular}
\caption{Signal-to-noise (SNR) of pairwise transverse velocity measurements from a joint analysis of CMB and galaxy surveys matching anticipated specifications of CMB-S4 and LSST for halos satisfying $M_h>5\times10^{13}M_\odot$. The first row corresponds to our  \baseline~consideration, following standard ILC-cleaning including all foregrounds and CMB-S4--like survey specifications. The second row corresponds to \unlensed~corresponds to omitting the cluster-lensing contribution to the CMB maps, otherwise identical to \baseline. Rows correspond to the same-colored data points on the left panels Fig.~\ref{fig:velo_estimate_pw_1}. The signal-to-noise for a given range of radial distance bins satisfy SNR$=\sqrt{\Delta\chi^2}=\sqrt{\chi_{\rm null}^2-\chi_{\rm bf}^2}$ where we set $v_{\langle ij\rangle}^{\rm model}=0$ ($=v_{\langle ij\rangle}^{\rm bf}$) when calculating $\chi^2_{\rm null}$ ($\chi_{\rm bf}^2$). Here $\tilde{v}_{\langle ij\rangle}$ is the measured pairwise velocity at the  and $v^{\rm bf}_{\langle ij\rangle}$ is the (best-fit) true pairwise-velocity signal calculated from simulations including only the ML effect. Defined this way, the SNR corresponds to chi-square significance of observations being better explained by the true pairwise-velocity signal as opposed to zero signal. Columns correspond to the three redshift bins we consider, matching the rows of Fig.~\ref{fig:velo_estimate_pw_1}. The first two of three sub-columns for each redshift correspond to taking only the data points with comoving distance between pairs satisfying $R<150$Mpc and $R>150$Mpc, respectively. The third sub-column of each 3-column corresponds to taking all data points. We only show SNR results if the $\Delta\chi^2_{\rm noML}=\chi_{\rm null}^2-\chi_{\rm noML}^2$ is negative, where we set $\tilde{v}^{\rm model}_{\langle ij\rangle}$ equal to $\tilde{v}^{\rm noML}_{\langle ij\rangle}$ when defining $\chi_{\rm noML}^2=\tilde{v}^{\rm noML}_{ij}$, and $\tilde{v}^{\rm noML}_{\langle ij\rangle}$ is the estimated pairwise transfer velocity in the absence of ML effect. The positivity of $\Delta\chi^2_{\rm noML}$ indices a non-zero significance of \textit{false} detection of pairwise transfer velocity `signal' in the absence of ML effect, i.e. detection of a bias.}\label{tab:SNR_1}
\vspace*{-0.5cm}
\end{table*}

\begin{table*}
\centering
\begin{tabular}{| r | c | c | c | c | c | c | c | c | c | c |}\hline
    SNR & 
       \multicolumn{3}{c}{$z\in[0.1,0.5]$} & 
       \multicolumn{3}{|c|}{$z\in[0.5,1.0]$} & 
       \multicolumn{3}{c|}{$z\in[1.0,2.0]$} & \\
       $R\in$ & $<150{\rm Mpc}$ & $>150{\rm Mpc}$ & All & $<150{\rm Mpc}$ & $>150{\rm Mpc}$ & All & $<150{\rm Mpc}$ & $>150{\rm Mpc}$ & All & Total \\ \hline 
       \baseline~& 5.11 & - & 5.11 &  3.04 & 1.41 & 3.6 & - & - & - & 6.3 \\
        {\unlensed} & 5.58 & 1.31 & 5.77 & 4.5 & 1.98 & 5.38 &  0.6 & 0.24 & 0.44 &  7.9 \\ \hline
\end{tabular}
\caption{The SNR of pairwise transverse velocity measurements from a joint analysis of CMB and galaxy surveys matching anticipated specifications of CMB-S4 and LSST for halos satisfying $M_h>10^{14}M_{\odot}$. Table otherwise identical to Table~\ref{tab:SNR_1}.\label{tab:SNR_2} Corresponding data points are shown in Fig.~\ref{fig:velo_estimate_pw_1}.} 
\vspace*{-0.5cm}
\end{table*}

Tables~\ref{tab:SNR_1}~and~\ref{tab:SNR_2} consist of three wide columns corresponding to our redshift ranges, with another three sub-columns for each, corresponding taking data points on smaller {comoving separations} $R<150$Mpc, larger ({$150{\rm Mpc}<R<200$Mpc}) and total ($R<200$Mpc, labelled `All') comoving distance separations. Empty (labelled `-') entries correspond to ${\Delta \chi^2_{\rm noML}}>0$. For \baseline~$M_h>5\times10^{13}M_\odot$, we find ${\rm SNR}\simeq5$ from $z\in[0.1,0.5]$, $\simeq7$ from $z\in[0.5,1.0]$ and $\simeq3$ from $z\in[1.0,2.0]$. In case the cluster lensing could be mitigated, we find for \unlensed~$M_h>5\times10^{13}M_\odot$, ${\rm SNR}\simeq10$ from $z\in[0.1,0.5]$, $\simeq9$ from $z\in[0.5,1,0]$ and $\sim4$ from $z\in[1.0,2,0]$. The total ${\rm SNR}$ is around 9 (14) for \baseline~(\unlensed). For a higher halo mass cut satisfying $M_h>10^{14}M_\odot$, we find it will be difficult to unambiguously detect the transverse velocity signal unless lensing could be mitigated. For the baseline results, we find ${\rm SNR}\simeq5\,(4)$ for the lowest and middle redshift bins with a total ${\rm SNR}\simeq6$. For the \unlensed~results, we find ${\rm SNR}\simeq6$ from $z\in[0.1,0.5]$, $\simeq6$ from $z\in[0.5,1,0]$ and $<1$ from $z\in[1.0,2,0]$ with total ${\rm SNR}\simeq8$. 

Figure~\ref{fig:velo_estimate_pw_1} also demonstrates the contributions of CIB and tSZ effects on the pairwise transfer-velocity signal on the middle panels of each row, with data in purple and pink color, respectively. We calculate these contributions by repeating our baseline analysis while omitting either CIB or tSZ foregrounds completely, and subtracting the results from the baseline results. This way we isolate and remove the net contribution from these frequency-dependent foregrounds while considering  the  ILC-cleaning procedure. For the halo masses and redshifts we consider in this analysis, we find CIB and tSZ foregrounds do not introduce significant biases (unlike what happened with that stacking analysis). Note however that in the stacking analysis we include halos of all redshifts and masses satisfying $M_h\gtrsim10^{12}$. As shown in \cite{Hotinli:2023ywh}, halos with lower masses and at higher redshifts lead to a more significant bias which considered a wider range of masses and redshifts, and observed the CIB and tSZ foregrounds can potentially introduce a significant bias for \textit{lower mass} halos (and in case of CIB, at higher redshifts). Nevertheless we find CIB and tSZ foregrounds contribute as significant confusion factors boosting the covariance, as can be seen from comparing the error bars on the left and middle panels of Fig.~\ref{fig:velo_estimate_pw_1}. 

We consider combination of CMB and LSS surveys matching SO and DESI specifications in Fig.~\ref{fig:numbercounts}. As the number of pairs scale like halo-number squared, around an order-of-magnitude lower number of halos for SO$\times$DESI compared to CMB-S4$\times$LSST, lead to $\sim1$M halo pairs for the pairwise transfer-velocity estimation. As a result we find the detection ${\rm SNR}$ such a combination of surveys is less than one for all mass and redshift ranges we consider, even in the absence of halo lensing. The dominant factor that contributes to the reduced detection significance is the halo number counts. We demonstrate this in Fig.~\ref{fig:numbercounts} for a range of galaxy number densities and a mass cut of $M_h<5\times10^{13}$. The top (bottom) rows of panels correspond to experimental specifications matching combination of DESI and SO (LSST and CMB-S4). The middle three panels we take a survey area matching LSST and CMB-S4, and vary the galaxy number count by a factor 1/2. The corresponding number of galaxy pairs satisfying $R<200$Mpc is $\sim18$M. For our \baseline~analysis we find ${\rm SNR}>1$ only on lowest panel while for \unlensed~we find surveys with galaxy number density only within a factor $\gtrsim0.2$ of LSST$\times$CMB-S4 can potentially reach ${\rm SNR}>1$.  

Last, we find our results are not sensitive to halo mass and photo-$z$ redshift errors considered in our matched filter defined in Sec.~\ref{sec:indv}, latter set equal to anticipated $\sigma_z=0.03(1+z)$ error from LSST, and we set the mass errors to 40 percent as anticipated by galaxy-galaxy-lensing cross correlation and SZ measurements~\citep[e.g.][]{Murata:2017zdo,Palmese:2019lkh,Ballardini:2019wxj}. Our results are shown in Fig.~\ref{fig:errors} for a range of mass and redshift error combinations. We find the ${\rm SNR}$ to be nearly identical between different considerations of error.

\begin{figure*}[t]
\centering
\includegraphics[width=1\linewidth]{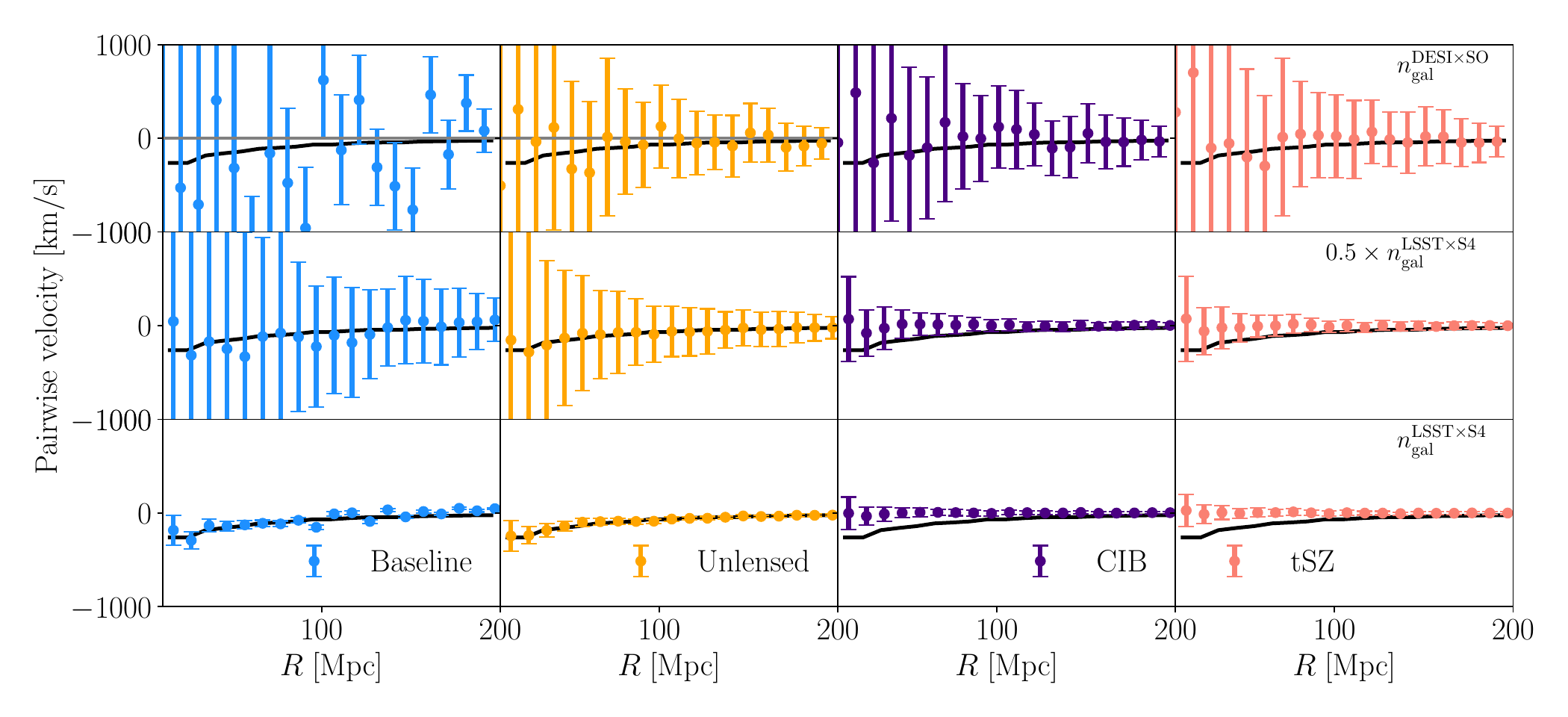}\vspace*{-0.3cm}
\caption{Pairwise velocity estimates for a range of galaxy survey specifications for mass cut of $M_h>5\times10^{13}$ and the lower redshift bin $z\in[0.1,0.5]$. The top panels correspond to a joint analysis of a DESI and SO like experiment. The following row of panels correspond to a joint analysis of CMB-S4 and LSST, with the exception that we reduce the galaxy density of LSST survey by a factor of 2. The lowest panels correspond to velocity estimates for our \baseline~choice (along with \unlensed, CIB and tSZ), identical to results the top row of panels in Fig.~\ref{fig:velo_estimate_pw_1}. We find the fidelity of the pairwise velocity measurement is significantly sensitive to the galaxy number count.} \label{fig:numbercounts}
\end{figure*}

\begin{figure*}[t]
\centering
\includegraphics[width=0.83\linewidth]{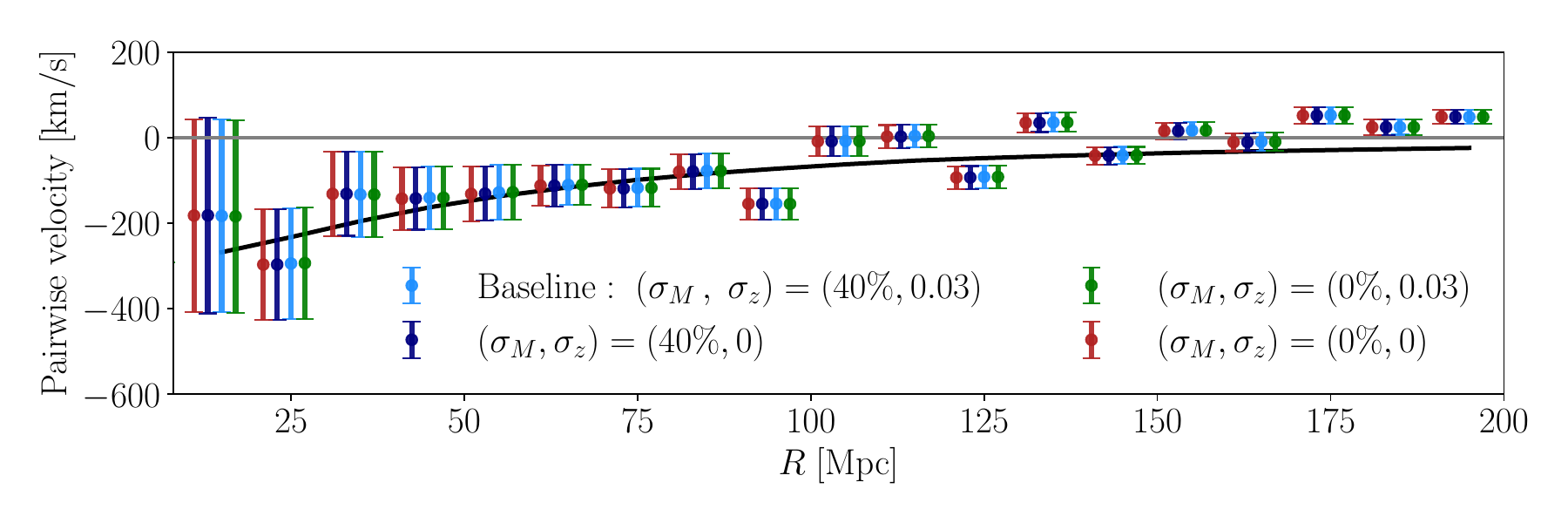}\vspace*{-0.5cm}\caption{The effect of halo mass and redshift errors on the measurement fidelity of the pairwise velocities for a combination of surveys matching LSST and CMB-S4 specifications. We compare perfect knowledge of halo redshifts and halo mass in our filters with 40 percent mass errors and the standard $\sigma_z=0.03(1+z)$ photo-$z$ error anticipated by the LSST survey. We find for these experimental specifications, the pairwise velocity estimation is insensitive to errors on halo mass and redshift in our filters. We discuss these result in Sec.~\ref{sec:results}.} \label{fig:errors}
\end{figure*}

\section{Discussion and conclusions}\label{sec:discussion}

Our analysis in Sec.~\ref{sec:detect} suggests Stage-4 CMB and LSS experiments will have the ability to detect the ML effect to high significance, in agreement with previous works~\citep[e.g.][]{Hotinli:2018yyc,Yasini:2018rrl,Hotinli:2020ntd,Hotinli:2021hih}. Here we included in the analysis for the first time simulations of extra-galactic foregrounds that are non-Gaussian and correlated with LSS, and found that they may be a significant limitation to an unambiguous detection of ML. 

The oriented stacking analysis showed that {some} foregrounds contribute a significant gradient to the final stacks, an effect we first described in \cite{Hotinli:2023ywh}, due to average motion of halos towards higher density regions in their LSS environment. As a result, CMB patches oriented along the transverse velocity of halos show an enhanced {signal} of CIB and tSZ foregrounds towards the velocity direction, leading to an aligned density gradient, in part mimicking the ML signal and biasing the results.} It could be possible, however, to mitigate this bias by applying mass and redshift cuts to the oriented stacking analysis. As shown in \cite{Hotinli:2023ywh}, such correlations between transverse velocities and matter gradient {are more enhanced for lower masses in general, and according to whether it is tSZ or CIB, it affects low or intermediate redshifts, respectively.} In an upcoming work we will further explore how applying high halo-mass cuts impact the contribution from these foregrounds to stacks, and what is the trade-off in terms of SNR between removing the most problematic objects and reducing the number of halos in the oriented-stacking analysis. 

Our results from the pairwise transverse-velocity estimation also demonstrate the significant impact of halo lensing on the ML detection, {which we find to be the main limiting factor}. The estimated pairwise-velocity signals shown in Figs.~\ref{fig:velo_estimate_pw_1}~and~\ref{fig:numbercounts} show how halo lensing boosts the error on the data points and derive measurements away from the predicted ML signal. A similar contribution is also apparent in oriented stacking, as the halo lensing leads to a significant gradient on the final stacks, dominating the contribution from black-body CMB, which can be seen in Fig.~\ref{fig:stacked_profiles1}. Note however that the contribution from halo lensing could in principle be mitigated by CMB delensing, the procedure of reversing the effects on lensing on the CMB, which has a variety of benefits for cosmological inference~\citep[see e.~g.][]{Hotinli:2021umk,Coulton:2019odk}. The coherence of unlensed CMB gradient at small scales makes the measurement of small-scale CMB lensing due to halos feasible~\cite[see e.g.][]{Schaan:2018tup,Horowitz:2017iql}. The methods developed for lensing reconstruction can in principle be applied for delensing. It is highly plausible that the adverse effect of halo lensing to ML detection can be mitigated effectively in the following years, and our analysis motivates further study on the prospects of delensing the small-scale CMB temperature maps. {Note also that the tSZ and CIB} bias we find in the stacking analysis is {not} apparent in our pairwise-transverse velocity estimation, {due to pairwise velocity estimation reducing the relevance of individual halo velocity direction.

In this paper we have only considered standard harmonic ILC cleaning method--which has been used for decades--to mitigate the contribution to CMB from frequency-dependent foregrounds~\citep{Tegmark:2003ve}. While standard ILC is the simplest method to reduce foregrounds, it is not always the most powerful one for the analysis of small-scale signatures. Recent years have seen an influx of more advanced methods such as the needled (wavelet) ILC~\citep{2009A&A...493..835D}, an implementation of ILC that works in the needlet (wavelet) domain. In this method, the input maps are decomposed into `needlets' at different angular scales and the ILC solution for the CMB is produced by minimizing the variance at each scale. This allows the procedure to depend on sky location and in principle be optimized against small-scale LSS sources, potentially improving the prospects of ML detection. 

{Another method for improving the measurement of black-body CMB signals is de-projection of frequency-dependent foregrounds such as the CIB from maps of observed CMB. We describe this method briefly in Appendix~\ref{app:deproj} and perform a preliminary application to our analysis. We find the improvement to the SNR from oriented stacking to be less than 10 percent. In what follows we will perform a more in-depth analysis to assess the prospects constrained ILC method by better modelling of the CIB foreground.}

Throughout our analysis we have implemented a matched filter based on the ensemble-averaged total CMB variance to minimize the contribution of foregrounds and primary (lensed) CMB to our stacks and pairwise velocity estimators. While proven optimal for previous spectra-level forecasts based on Gaussian approximation of foregrounds~\citep[see e.g.][]{Hotinli:2020ntd}, it is conceivable that our filters can be significantly improved in multitude of ways, taking into account halo-specific foreground profiles. First, as the halo-induced non-Gaussian foregrounds dominate the contributions to the oriented stacks, we could in principle use filters that aim to minimize the contribution from foregrounds in a way depending on the spatial profile of the foreground, the halo mass and redshift; rather than weighting-down profiles against ensemble-average foreground variance. Note, however, that the dominant source of confusion for our stacking analysis is the matter-gradients correlated with transverse velocities; the profile of the contributions depend on the overall density gradient of the halo environment, in addition to halo mass and redshift. As a results, the optimal filters that would mitigate such contributions should take into account the full contribution to the stacks as discussed in~\citep{Hotinli:2023ywh}, which could be difficult to model. Furthermore, in our pairwise transverse velocity analysis, we have chosen to individually filter the pairs of CMB patches {to detect individual halo velocities and then combine these to compute the pairwise velocities.} In principle one can develop filters optimized for {the extraction of pairwise velocities directly from the CMB map}. We leave a detailed study on developing more optimal filtering techniques for the purpose of ML detection to an upcoming work.    

Note we have omitted forecasting for planned Stage-5 surveys such as CMB-HD~\citep{CMB-HD:2022bsz, Sehgal:2020yja, Sehgal:2019ewc}, a futuristic follow-up of CMB-S4 with CMB white noise RMS anticipated to reach $\Delta_T\simeq0.2\mu{\rm K}'$ and beam $\sim15''$, or MegaMapper~\citep{Schlegel:2019eqc}, the spectroscopic follow-up of LSST. Such surveys will have access to smaller scales compared to what we model here, and a forecast that would be representative of the precision of these surveys require improved simulations with higher-resolution maps, which are currently not widely available. 

Overall our analysis shows that upcoming Stage-4 surveys like CMB-S4 and LSST can detect the ML signal within $10\!-\!20\sigma$ significance, while the prospects of detection for ongoing Stage-3 surveys such as DESI and SO are likely less optimistic due to lower number of halos. In order to maximize the detection prospects with oriented stacking, the CIB and tSZ foregrounds will need to be mitigated more effectively as compared to standard ILC-cleaning, or modelled{,} and model parameters marginalized with informed priors. Halo lensing likely will adversely impact the detection prospects in case not accounted for, in particular for the pairwise-velocity estimation. We find large halo masses $M_h>5\times10^{13}M_\odot$ at lower redshifts $z\lesssim1$ will likely provide the bulk of the detection SNR. Although potentially difficult, we nevertheless anticipate the exciting  prospect of detecting ML effect in the near future warranting continuation of dedicated work to overcome the challenges we highlighted here. 

\vspace*{-0.4cm}

\section{Acknowledgements}

We thank Sanjaykumar Patil, Fiona McCarthy, Marcelo Alvarez, Reijo Keskitalo, Simone Ferraro, Kendrick Smith, Matthew Jonhson for useful discussions. We Sanjaykumar Patil and Nareg Mirzatuny  for collaboration at the early stages of this work. SCH  is supported by the P.~J.~E.~Peebles Fellowship at Perimeter Institute for Theoretical Physics. This research was supported in part by Perimeter Institute for Theoretical Physics. Research at Perimeter Institute is supported by the Government of Canada through the Department of Innovation, Science and Economic Development Canada and by the Province of Ontario through the Ministry of Research, Innovation and Science. EP is supported by NASA grant 80NSSC23K0747.
This work was  performed in part at Aspen Center for Physics, which is supported by National Science Foundation grant PHY-2210452.
The authors acknowledge the Center for Advanced Research Computing (CARC) at the University of Southern California for providing computing resources that have contributed to the research results reported within this publication. This work was also carried out at the Advanced Research Computing at Hopkins (ARCH) core facility  (rockfish.jhu.edu), which is supported by the National Science Foundation (NSF) grant number OAC1920103. SCH was in part supported by the Horizon Fellowship from Johns Hopkins University. This research used resources of the National Energy Research Scientific Computing Center (NERSC), a U.S. Department of Energy Office of Science User Facility located at Lawrence Berkeley National Laboratory, operated under Contract No. DE-AC02-05CH11231 using NERSC award HEP-ERCAPmp107. This research was supported in part by grant NSF PHY-1748958 to the Kavli Institute for Theoretical Physics (KITP).

\section{Appendix}

\subsection{ILC cleaning prescription}\label{sec:ILC-appendix}

The distinct frequency dependence of black-body signals such as CIB, tSZ and point sources allows separating these foregrounds from black-body signals such as primary CMB and the ML effect. Component separation strategies aim at reconstructing the best possible map that primarily contain black-body signals.  
The most common ``component separation" strategy is an internal linear combination (ILC) method which aims at reconstructing the minimum variance CMB map. 
In this work we use a standard harmonic-space ILC procedure as prescribed in~\citep{Tegmark:2003ve}, which we detail below. We describe the experimental specifications we use for the CMB instrumental noise in Table~\ref{tab:cmb_specs}.

We write the covariance between the de-beamed CMB at different frequencies as a matrix
\be
\boldsymbol{C}_\ell=C_\ell^{TT} \boldsymbol{e}\boldsymbol{e}^\dagger+\boldsymbol{C}_\ell^{\rm XG}+(\boldsymbol{B}^{-1}\boldsymbol{N})_\ell\,,
\ee
where $C_\ell^{TT}$ contains the black-body component of the CMB (primary lensed CMB, kSZ and the ML effects), $\boldsymbol{e}=\{1,1,1\,\ldots\}$, $\boldsymbol{C}_\ell^{\rm XG}$ contains the correlated CIB, tSZ and radio point sources, and $(\boldsymbol{B}^{-1}\boldsymbol{N})_\ell$ is the de-beamed instrumental noise covariance (which we assume diagonal). The blackbody component after the ILC procedure satisfy
\be
\hat{\Theta}_{\ell m}=\boldsymbol{w}_\ell^{\dagger}\boldsymbol{\Theta}_{\ell m}\,,
\ee
where the weights that minimize the variance of the multipole moments $\hat{\Theta}_{\ell m}$ are given by
\be
\boldsymbol{w}_\ell=\frac{(\boldsymbol{C}_\ell)^{-1}\boldsymbol{e}}{\boldsymbol{e}^\dagger(\boldsymbol{C}_\ell)^{-1}\boldsymbol{e}}\,.
\ee

\subsection{CIB de-projection}\label{app:deproj}

{One method to improve the measurements of the black-body CMB signals is the deprojection of various frequency-dependent foregrounds such as CIB to minimize their contamination, also known as `constrained ILC'~\citep[see for a recent application, for example][]{McCarthy:2023hpa}. Since CIB contribute significantly to the oriented stacking analysis, constrained ILC can be a valuable method to enhance detection prospects of ML effect in the future. Note however that these methods require knowledge of the frequency-dependence of the foregrounds that are deprojected, which does not always have a well-understood spectral energy distribution (SED) as in the case of CIB, for example. 

As a preliminary analysis, we use the publicly available \texttt{pyilc}\footnote{Available at \hyperlink{https://github.com/jcolinhill/pyilc}{jcolinhill/pyilc} and see~\cite{McCarthy:2023hpa,McCarthy:2023cwg,2011MNRAS.410.2481R,2009ApJ...694..222C}.} code to deproject the CIB foreground from our \texttt{websky} maps using harmonic constrained ILC. We model the frequency dependence of CIB with $\nu^\beta$ frequency dependence and effective dust temperature $T_d=T_0(1+z)^\alpha$ setting $\beta=0.2$, $\alpha=0.2$ and $T_0=20.7$ following~\citep{Stein:2018lrh}. While we find CIB deprojection reduces the contribution to CMB by around a factor $\sim\,2$ on $\sim$\,arcminute scales, we find the effect on the final ML detection SNR from oriented stacking to be {less than 10 percent}.}

\subsection{Stacking}\label{sec:stacking-appendix}

For a given pixel in the sky, we define a $3\times3$ rotation matrix pivoted at the location of the halo (labelled $i$ below) as
\be
\boldsymbol{M}_{\rm rot}^i=\begin{bmatrix}
\,\,\,\,
\mathcal{I}_1\cdot\mathcal{I}_{v1}^i\,\,\,\, & \,\,\,\,\mathcal{I}_1\cdot\mathcal{I}_{v2}^i\,\,\,\, & \,\,\,\,\mathcal{I}_1\cdot\mathcal{I}_{v3}^i\,\,\,\,\\
\mathcal{I}_2\cdot\mathcal{I}_{v1}^i & \mathcal{I}_2\cdot\mathcal{I}_{v2}^i & \mathcal{I}_2\cdot\mathcal{I}_{v3}^i\\
\mathcal{I}_3\cdot\mathcal{I}_{v1}^i & \mathcal{I}_3\cdot\mathcal{I}_{v2}^i & \mathcal{I}_3\cdot\mathcal{I}_{v3}^i\\
\end{bmatrix}
\ee
where 
\be 
\mathcal{I}_1\equiv\{1,0,0\} ,\,~\mathcal{I}_2\equiv\{0,1,0\}, \,~\mathcal{I}_3\equiv\{0,0,1\}
\ee
and
\be
\mathcal{I}_{v1}^i&&\equiv\left\{ \hat{x}^i, \hat{y}^i, \hat{z}^i\right\}\\
\mathcal{I}_{v2}^i &&\equiv \left\{ \hat{v}_{\perp x}^i, \hat{v}_{\perp y}^i, \hat{v}_{\perp z}^i \right\}\\
\mathcal{I}_{v3}^i && \equiv 
\left\{ 
\hat{y}^i\,\hat{v}_{\perp z}^i- \hat{z}^i\,\hat{v}_{\perp y}^i,\,
\hat{z}^i\,\hat{v}_{\perp x}^i- \hat{x}^i\,\hat{v}_{\perp z}^i,\,
\hat{x}^i\,\hat{v}_{\perp y}^i- \hat{y}^i\,\hat{v}_{\perp x}^i
\right\}\,.
\ee
Here, $\hat{x}^i=x^i/\chi$, $\hat{y}^i=y^i/\chi$, $\hat{z}^i=z^i/\chi$ where $x^i,y^i,z^i$ are the comoving 3-dimensional coordinates of the halo and \be(\chi^i)^2=(x^i)^2+(y^i)^2+(z^i)^2\ee where $\chi^i$ is the comoving distance to the halo. The velocity unit vectors satisfy 
\be
\hat{v}_{\perp x}^i=\frac{\boldsymbol{v}_\perp^i\cdot\hat{\boldsymbol{x}}^i}{|\boldsymbol{v}_\perp^i|},\,\,\, \hat{v}_{\perp y}^i=\frac{\boldsymbol{v}_\perp^i\cdot\hat{\boldsymbol{y}}^i}{|\boldsymbol{v}_\perp^i|}\,\,\,{\rm and}\,\,\,
\hat{v}_{\perp z}^i=\frac{\boldsymbol{v}_\perp^i\cdot\hat{\boldsymbol{z}}^i}{|\boldsymbol{v}_\perp^i|}
\ee 
where $\boldsymbol{v}_\perp^i$ is the three-dimensional bulk transverse velocity at the location of the halo and $\hat{\boldsymbol{x}}^i$, $\hat{\boldsymbol{y}}^i$, $\hat{\boldsymbol{z}}^i$ are the three-dimensional Cartesian unit vectors corresponding to the pixel's location on the 2-sphere. The Cartesian coordinates for a \textit{pixel} on the \textit{rotated} map then satisfies 
\be
\begin{bmatrix}
x_{\rm rot}\\
y_{\rm rot}\\
z_{\rm rot}
\end{bmatrix}=\boldsymbol{M}_{\rm tr}^i
\begin{bmatrix}
x_{\rm map}\\
y_{\rm map}\\
z_{\rm map}
\end{bmatrix}\,,\label{eq:stackingrot}
\ee
where $x_{\rm map},y_{\rm map},z_{\rm map}$ are the Cartesian coordinates of the pixel on the \textit{input} patch prior to rotation, and $x_{\rm rot},y_{\rm rot},z_{\rm rot}$ are the Cartesian coordinates of the pixel on the rotated map. We choose $x_{\rm map},y_{\rm map},z_{\rm map}$ to correspond to $N_p\times N_p$ pixels covering the range $[-\lambda r/r_s,,+\lambda r/r_s]$ in two orthogonal directions on the 2-sphere and choose $\lambda=2.5$ and $N_p=21$ unless we state otherwise. 

\bibliography{main}

\begin{thebibliography}{}
\expandafter\ifx\csname natexlab\endcsname\relax\def\natexlab#1{#1}\fi
\providecommand{\url}[1]{\href{#1}{#1}}
\providecommand{\dodoi}[1]{doi:~\href{http://doi.org/#1}{\nolinkurl{#1}}}
\providecommand{\doeprint}[1]{\href{http://ascl.net/#1}{\nolinkurl{http://ascl.net/#1}}}
\providecommand{\doarXiv}[1]{\href{https://arxiv.org/abs/#1}{\nolinkurl{https://arxiv.org/abs/#1}}}

\bibitem[{Abazajian {et~al.}(2022)}]{CMB-S4:2022ght}
Abazajian, K., {et~al.} 2022.
\newblock \doarXiv{2203.08024}

\bibitem[{{Abazajian} {et~al.}(2016)}]{Abazajian:2016yjj}
{Abazajian}, K.~N., {et~al.} 2016, arXiv e-prints, arXiv:1610.02743.
\newblock \doarXiv{1610.02743}

\bibitem[{{Ade} {et~al.}(2019){Ade}, {Aguirre}, {Ahmed}, {Aiola}, {Ali},
  {Alonso}, {Alvarez}, {Arnold}, {Ashton}, {Austermann}, \&
  et~al.}]{Ade:2018sbj}
{Ade}, P., {Aguirre}, J., {Ahmed}, Z., {et~al.} 2019, \jcap, 2019, 056,
  \dodoi{10.1088/1475-7516/2019/02/056}

\bibitem[{Ade {et~al.}(2016)}]{Planck:2015ywj}
Ade, P. A.~R., {et~al.} 2016, Astron. Astrophys., 586, A140,
  \dodoi{10.1051/0004-6361/201526328}

\bibitem[{Aghanim {et~al.}(2020)}]{Planck:2018vyg}
Aghanim, N., {et~al.} 2020, Astron. Astrophys., 641, A6,
  \dodoi{10.1051/0004-6361/201833910}

\bibitem[{Aiola {et~al.}(2022)}]{CMB-HD:2022bsz}
Aiola, S., {et~al.} 2022.
\newblock \doarXiv{2203.05728}

\bibitem[{Anil~Kumar {et~al.}(2022)Anil~Kumar, Sato-Polito, Kamionkowski, \&
  Hotinli}]{AnilKumar:2022flx}
Anil~Kumar, N., Sato-Polito, G., Kamionkowski, M., \& Hotinli, S.~C. 2022,
  Phys. Rev. D, 106, 063533, \dodoi{10.1103/PhysRevD.106.063533}

\bibitem[{Ballardini {et~al.}(2019)Ballardini, Matthewson, \&
  Maartens}]{Ballardini:2019wxj}
Ballardini, M., Matthewson, W.~L., \& Maartens, R. 2019, Mon. Not. Roy. Astron.
  Soc., 489, 1950, \dodoi{10.1093/mnras/stz2258}

\bibitem[{Bayer {et~al.}(2023)Bayer, Modi, \& Ferraro}]{Bayer:2022vid}
Bayer, A.~E., Modi, C., \& Ferraro, S. 2023, JCAP, 06, 046,
  \dodoi{10.1088/1475-7516/2023/06/046}

\bibitem[{{Birkinshaw} \& {Gull}(1983)}]{1983Natur.302..315B}
{Birkinshaw}, M., \& {Gull}, S.~F. 1983, \nat, 302, 315,
  \dodoi{10.1038/302315a0}

\bibitem[{Cayuso {et~al.}(2023)Cayuso, Bloch, Hotinli, Johnson, \&
  McCarthy}]{Cayuso:2021ljq}
Cayuso, J., Bloch, R., Hotinli, S.~C., Johnson, M.~C., \& McCarthy, F. 2023,
  JCAP, 02, 051, \dodoi{10.1088/1475-7516/2023/02/051}

\bibitem[{{Chen} \& {Wright}(2009)}]{2009ApJ...694..222C}
{Chen}, X., \& {Wright}, E.~L. 2009, \apj, 694, 222,
  \dodoi{10.1088/0004-637X/694/1/222}

\bibitem[{Coulton {et~al.}(2022)Coulton, Feldman, Maamari, Pierpaoli, Yasini,
  \& Dolag}]{Coulton:2021ekh}
Coulton, W.~R., Feldman, S., Maamari, K., {et~al.} 2022, Mon. Not. Roy. Astron.
  Soc., 513, 2252, \dodoi{10.1093/mnras/stac1017}

\bibitem[{Coulton {et~al.}(2020)Coulton, Meerburg, Baker, Hotinli,
  Duivenvoorden, \& van Engelen}]{Coulton:2019odk}
Coulton, W.~R., Meerburg, P.~D., Baker, D.~G., {et~al.} 2020, Phys. Rev. D,
  101, 123504, \dodoi{10.1103/PhysRevD.101.123504}

\bibitem[{De~Bernardis {et~al.}(2017)}]{DeBernardis:2016pdv}
De~Bernardis, F., {et~al.} 2017, JCAP, 03, 008,
  \dodoi{10.1088/1475-7516/2017/03/008}

\bibitem[{{Delabrouille} {et~al.}(2009){Delabrouille}, {Cardoso}, {Le Jeune},
  {Betoule}, {Fay}, \& {Guilloux}}]{2009A&A...493..835D}
{Delabrouille}, J., {Cardoso}, J.~F., {Le Jeune}, M., {et~al.} 2009, \aap, 493,
  835, \dodoi{10.1051/0004-6361:200810514}

\bibitem[{{DESI Collaboration} {et~al.}(2016){DESI Collaboration}, {Aghamousa},
  {Aguilar}, {Ahlen}, {Alam}, {Allen}, {Allende Prieto}, {Annis}, {Bailey},
  {Balland}, \& et~al.}]{DESI:2016fyo}
{DESI Collaboration}, {Aghamousa}, A., {Aguilar}, J., {et~al.} 2016, arXiv
  e-prints, arXiv:1611.00036.
\newblock \doarXiv{1611.00036}

\bibitem[{Escoffier {et~al.}(2016)Escoffier, Cousinou, Tilquin, Pisani,
  Aguichine, de~la Torre, Ealet, Gillard, \& Jullo}]{Escoffier:2016qnf}
Escoffier, S., Cousinou, M.~C., Tilquin, A., {et~al.} 2016.
\newblock \doarXiv{1606.00233}

\bibitem[{Ferraro {et~al.}(2022)Ferraro, Schaan, \&
  Pierpaoli}]{Ferraro:2022twg}
Ferraro, S., Schaan, E., \& Pierpaoli, E. 2022.
\newblock \doarXiv{2205.10332}

\bibitem[{Ferreira {et~al.}(1999)Ferreira, Juszkiewicz, Feldman, Davis, \&
  Jaffe}]{Ferreira:1998id}
Ferreira, P.~G., Juszkiewicz, R., Feldman, H.~A., Davis, M., \& Jaffe, A.~H.
  1999, Astrophys. J. Lett., 515, L1, \dodoi{10.1086/311959}

\bibitem[{Guachalla {et~al.}(2023)Guachalla, Schaan, Hadzhiyska, \&
  Ferraro}]{Guachalla:2023lbx}
Guachalla, B.~R., Schaan, E., Hadzhiyska, B., \& Ferraro, S. 2023.
\newblock \doarXiv{2312.12435}

\bibitem[{{Gurvits} \& {Mitrofanov}(1986)}]{Gurvitz:1986ab}
{Gurvits}, L.~I., \& {Mitrofanov}, I.~G. 1986, \nat, 324, 349,
  \dodoi{10.1038/324349a0}

\bibitem[{Hadzhiyska {et~al.}(2023)Hadzhiyska, Ferraro, Guachalla, \&
  Schaan}]{Hadzhiyska:2023nig}
Hadzhiyska, B., Ferraro, S., Guachalla, B.~R., \& Schaan, E. 2023.
\newblock \doarXiv{2312.12434}

\bibitem[{Hall \& Challinor(2014)}]{Hall:2014wna}
Hall, A., \& Challinor, A. 2014, Phys. Rev., D90, 063518,
  \dodoi{10.1103/PhysRevD.90.063518}

\bibitem[{{Hand} {et~al.}(2012)}]{2012PhRvL.109d1101H}
{Hand}, N., {et~al.} 2012, \prl, 109, 041101,
  \dodoi{10.1103/PhysRevLett.109.041101}

\bibitem[{Horowitz {et~al.}(2019)Horowitz, Ferraro, \&
  Sherwin}]{Horowitz:2017iql}
Horowitz, B., Ferraro, S., \& Sherwin, B.~D. 2019, Mon. Not. Roy. Astron. Soc.,
  485, 3919, \dodoi{10.1093/mnras/stz566}

\bibitem[{Hotinli {et~al.}(2023{\natexlab{a}})Hotinli, Ferraro, Holder,
  Johnson, Kamionkowski, \& La~Plante}]{Hotinli:2022jna}
Hotinli, S.~C., Ferraro, S., Holder, G.~P., {et~al.} 2023{\natexlab{a}}, Phys.
  Rev. D, 107, 103517, \dodoi{10.1103/PhysRevD.107.103517}

\bibitem[{Hotinli {et~al.}(2022{\natexlab{a}})Hotinli, Holder, Johnson, \&
  Kamionkowski}]{Hotinli:2022wbk}
Hotinli, S.~C., Holder, G.~P., Johnson, M.~C., \& Kamionkowski, M.
  2022{\natexlab{a}}, JCAP, 10, 026, \dodoi{10.1088/1475-7516/2022/10/026}

\bibitem[{{Hotinli} {et~al.}(2021){Hotinli}, {Johnson}, \&
  {Meyers}}]{Hotinli:2020ntd}
{Hotinli}, S.~C., {Johnson}, M.~C., \& {Meyers}, J. 2021, \prd, 103, 043536,
  \dodoi{10.1103/PhysRevD.103.043536}

\bibitem[{Hotinli {et~al.}(2019{\natexlab{a}})Hotinli, Mertens, Johnson, \&
  Kamionkowski}]{Hotinli:2019wdp}
Hotinli, S.~C., Mertens, J.~B., Johnson, M.~C., \& Kamionkowski, M.
  2019{\natexlab{a}}, Phys. Rev. D, 100, 103528,
  \dodoi{10.1103/PhysRevD.100.103528}

\bibitem[{Hotinli {et~al.}(2022{\natexlab{b}})Hotinli, Meyers, Trendafilova,
  Green, \& van Engelen}]{Hotinli:2021umk}
Hotinli, S.~C., Meyers, J., Trendafilova, C., Green, D., \& van Engelen, A.
  2022{\natexlab{b}}, JCAP, 04, 020, \dodoi{10.1088/1475-7516/2022/04/020}

\bibitem[{Hotinli {et~al.}(2023{\natexlab{b}})Hotinli, Pierpaoli, Ferraro, \&
  Smith}]{Hotinli:2023ywh}
Hotinli, S.~C., Pierpaoli, E., Ferraro, S., \& Smith, K. 2023{\natexlab{b}},
  Phys. Rev. D, 108, 083508, \dodoi{10.1103/PhysRevD.108.083508}

\bibitem[{Hotinli {et~al.}(2021)Hotinli, Smith, Madhavacheril, \&
  Kamionkowski}]{Hotinli:2021hih}
Hotinli, S.~C., Smith, K.~M., Madhavacheril, M.~S., \& Kamionkowski, M. 2021,
  Phys. Rev. D, 104, 083529, \dodoi{10.1103/PhysRevD.104.083529}

\bibitem[{Hotinli {et~al.}(2019{\natexlab{b}})Hotinli, Meyers, Dalal, Jaffe,
  Johnson, Mertens, Münchmeyer, Smith, \& van Engelen}]{Hotinli:2018yyc}
Hotinli, S.~C., Meyers, J., Dalal, N., {et~al.} 2019{\natexlab{b}}, Phys. Rev.
  Lett., 123, 061301, \dodoi{10.1103/PhysRevLett.123.061301}

\bibitem[{{Leauthaud} {et~al.}(2011){Leauthaud}, {Tinker}, {Behroozi}, {Busha},
  \& {Wechsler}}]{2011ApJ...738...45L}
{Leauthaud}, A., {Tinker}, J., {Behroozi}, P.~S., {Busha}, M.~T., \&
  {Wechsler}, R.~H. 2011, \apj, 738, 45, \dodoi{10.1088/0004-637X/738/1/45}

\bibitem[{{Lee} {et~al.}(2019){Lee}, {Abitbol}, {Adachi}, {Ade}, {Aguirre},
  {Ahmed}, {Aiola}, {Ali}, {Alonso}, {Alvarez}, \& et~al.}]{Abitbol:2019nhf}
{Lee}, A., {Abitbol}, M.~H., {Adachi}, S., {et~al.} 2019, in Bulletin of the
  American Astronomical Society, Vol.~51, 147.
\newblock \doarXiv{1907.08284}

\bibitem[{{Lewis} \& {Challinor}(2006)}]{Lewis:2006fu}
{Lewis}, A., \& {Challinor}, A. 2006, \physrep, 429, 1,
  \dodoi{10.1016/j.physrep.2006.03.002}

\bibitem[{Li {et~al.}(2022)Li, Puglisi, Madhavacheril, \& Alvarez}]{Li:2021ial}
Li, Z., Puglisi, G., Madhavacheril, M.~S., \& Alvarez, M.~A. 2022, JCAP, 08,
  029, \dodoi{10.1088/1475-7516/2022/08/029}

\bibitem[{{LSST Science Collaboration} {et~al.}(2009){LSST Science
  Collaboration}, {Abell}, {Allison}, {Anderson}, {Andrew}, {Angel}, {Armus},
  {Arnett}, {Asztalos}, {Axelrod}, \& et~al.}]{2009arXiv0912.0201L}
{LSST Science Collaboration}, {Abell}, P.~A., {Allison}, J., {et~al.} 2009,
  ArXiv e-prints.
\newblock \doarXiv{0912.0201}

\bibitem[{{Madhavacheril} {et~al.}(2019){Madhavacheril}, {Battaglia}, {Smith},
  \& {Sievers}}]{Madhavacheril:2019buy}
{Madhavacheril}, M.~S., {Battaglia}, N., {Smith}, K.~M., \& {Sievers}, J.~L.
  2019, arXiv e-prints, arXiv:1901.02418.
\newblock \doarXiv{1901.02418}

\bibitem[{McCarthy \& Hill(2023{\natexlab{a}})}]{McCarthy:2023hpa}
McCarthy, F., \& Hill, J.~C. 2023{\natexlab{a}}.
\newblock \doarXiv{2307.01043}

\bibitem[{McCarthy \& Hill(2023{\natexlab{b}})}]{McCarthy:2023cwg}
---. 2023{\natexlab{b}}.
\newblock \doarXiv{2308.16260}

\bibitem[{{M{\"u}nchmeyer} {et~al.}(2019){M{\"u}nchmeyer}, {Madhavacheril},
  {Ferraro}, {Johnson}, \& {Smith}}]{Munchmeyer:2018eey}
{M{\"u}nchmeyer}, M., {Madhavacheril}, M.~S., {Ferraro}, S., {Johnson}, M.~C.,
  \& {Smith}, K.~M. 2019, \prd, 100, 083508,
  \dodoi{10.1103/PhysRevD.100.083508}

\bibitem[{Murata {et~al.}(2018)Murata, Nishimichi, Takada, Miyatake, Shirasaki,
  More, Takahashi, \& Osato}]{Murata:2017zdo}
Murata, R., Nishimichi, T., Takada, M., {et~al.} 2018, Astrophys. J., 854, 120,
  \dodoi{10.3847/1538-4357/aaaab8}

\bibitem[{Palmese {et~al.}(2020)}]{Palmese:2019lkh}
Palmese, A., {et~al.} 2020, Mon. Not. Roy. Astron. Soc., 493, 4591,
  \dodoi{10.1093/mnras/staa526}

\bibitem[{{Remazeilles} {et~al.}(2011){Remazeilles}, {Delabrouille}, \&
  {Cardoso}}]{2011MNRAS.410.2481R}
{Remazeilles}, M., {Delabrouille}, J., \& {Cardoso}, J.-F. 2011, \mnras, 410,
  2481, \dodoi{10.1111/j.1365-2966.2010.17624.x}

\bibitem[{{Sachs} \& {Wolfe}(1967)}]{1967ApJ...147...73S}
{Sachs}, R.~K., \& {Wolfe}, A.~M. 1967, \apj, 147, 73, \dodoi{10.1086/148982}

\bibitem[{{Sazonov} \& {Sunyaev}(1999)}]{Sazonov:1999zp}
{Sazonov}, S.~Y., \& {Sunyaev}, R.~A. 1999, \mnras, 310, 765,
  \dodoi{10.1046/j.1365-8711.1999.02981.x}

\bibitem[{Schaan \& Ferraro(2019)}]{Schaan:2018tup}
Schaan, E., \& Ferraro, S. 2019, Phys. Rev. Lett., 122, 181301,
  \dodoi{10.1103/PhysRevLett.122.181301}

\bibitem[{Schaan {et~al.}(2017)Schaan, Krause, Eifler, Dor\'e, Miyatake,
  Rhodes, \& Spergel}]{Schaan:2016ois}
Schaan, E., Krause, E., Eifler, T., {et~al.} 2017, Phys. Rev. D, 95, 123512,
  \dodoi{10.1103/PhysRevD.95.123512}

\bibitem[{Schaan {et~al.}(2016)}]{ACTPol:2015teu}
Schaan, E., {et~al.} 2016, Phys. Rev. D, 93, 082002,
  \dodoi{10.1103/PhysRevD.93.082002}

\bibitem[{Schlegel {et~al.}(2019)}]{Schlegel:2019eqc}
Schlegel, D.~J., {et~al.} 2019.
\newblock \doarXiv{1907.11171}

\bibitem[{Sehgal {et~al.}(2019)}]{Sehgal:2019ewc}
Sehgal, N., {et~al.} 2019.
\newblock \doarXiv{1906.10134}

\bibitem[{Sehgal {et~al.}(2020)}]{Sehgal:2020yja}
---. 2020.
\newblock \doarXiv{2002.12714}

\bibitem[{{Smith} {et~al.}(2018){Smith}, {Madhavacheril}, {M{\"u}nchmeyer},
  {Ferraro}, {Giri}, \& {Johnson}}]{Smith:2018bpn}
{Smith}, K.~M., {Madhavacheril}, M.~S., {M{\"u}nchmeyer}, M., {et~al.} 2018,
  arXiv e-prints, arXiv:1810.13423.
\newblock \doarXiv{1810.13423}

\bibitem[{Soergel {et~al.}(2016)}]{DES:2016umt}
Soergel, B., {et~al.} 2016, Mon. Not. Roy. Astron. Soc., 461, 3172,
  \dodoi{10.1093/mnras/stw1455}

\bibitem[{Stein {et~al.}(2019)Stein, Alvarez, \& Bond}]{Stein:2018lrh}
Stein, G., Alvarez, M.~A., \& Bond, J.~R. 2019, Mon. Not. Roy. Astron. Soc.,
  483, 2236, \dodoi{10.1093/mnras/sty3226}

\bibitem[{Stein {et~al.}(2020)Stein, Alvarez, Bond, van Engelen, \&
  Battaglia}]{Stein:2020its}
Stein, G., Alvarez, M.~A., Bond, J.~R., van Engelen, A., \& Battaglia, N. 2020.
\newblock \doarXiv{2001.08787}

\bibitem[{{Sunyaev} \& {Zeldovich}(1980)}]{1980ARA&A..18..537S}
{Sunyaev}, R.~A., \& {Zeldovich}, I.~B. 1980, \araa, 18, 537,
  \dodoi{10.1146/annurev.aa.18.090180.002541}

\bibitem[{{Sunyaev} \& {Zeldovich}(1972)}]{1972CoASP...4..173S}
{Sunyaev}, R.~A., \& {Zeldovich}, Y.~B. 1972, Comments on Astrophysics and
  Space Physics, 4, 173

\bibitem[{Tegmark {et~al.}(2003)Tegmark, de~Oliveira-Costa, \&
  Hamilton}]{Tegmark:2003ve}
Tegmark, M., de~Oliveira-Costa, A., \& Hamilton, A. 2003, Phys. Rev. D, 68,
  123523, \dodoi{10.1103/PhysRevD.68.123523}

\bibitem[{{The LSST Dark Energy Science Collaboration} {et~al.}(2018){The LSST
  Dark Energy Science Collaboration}, {Mandelbaum}, {Eifler}, {Hlo{\v{z}}ek},
  {Collett}, {Gawiser}, \& {Scolnic}}]{LSSTDarkEnergyScience:2018jkl}
{The LSST Dark Energy Science Collaboration}, {Mandelbaum}, R., {Eifler}, T.,
  {et~al.} 2018, arXiv e-prints, arXiv:1809.01669.
\newblock \doarXiv{1809.01669}

\bibitem[{Yasini {et~al.}(2020)Yasini, Alvarez, Schaan, Maamari, s.~Mazinani,
  Mirzatuny, \& Pierpaoli}]{Yasini2020}
Yasini, S., Alvarez, M., Schaan, E., {et~al.} 2020, Journal of Open Source
  Software, 5, 2608, \dodoi{10.21105/joss.02608}

\bibitem[{{Yasini} {et~al.}(2019){Yasini}, {Mirzatuny}, \&
  {Pierpaoli}}]{Yasini:2018rrl}
{Yasini}, S., {Mirzatuny}, N., \& {Pierpaoli}, E. 2019, \apjl, 873, L23,
  \dodoi{10.3847/2041-8213/ab0bfe}

\bibitem[{{Zel'Dovich}(1970)}]{1970A&A.....5...84Z}
{Zel'Dovich}, Y.~B. 1970, \aap, 500, 13

\bibitem[{{Zeldovich} \& {Sunyaev}(1969)}]{1969Ap&SS...4..301Z}
{Zeldovich}, Y.~B., \& {Sunyaev}, R.~A. 1969, \apj, 4, 301,
  \dodoi{10.1007/BF00661821}

\end{thebibliography}

\end{document}